\documentclass[11pt]{article}

\newcommand{\cut}[1]{}
\usepackage{xfrac}
\usepackage{graphicx}
\usepackage{amssymb}        
\usepackage{hyperref}
\usepackage{amsfonts}
\usepackage{amsmath}
\usepackage{verbatim}
\usepackage[font=footnotesize]{caption}
\usepackage{subcaption}
\usepackage{blindtext}
\usepackage{booktabs}
\usepackage{authblk}

\begin{document}
\captionsetup{justification=justified}
\captionsetup{width=.8\linewidth}

\title{Global warming: Temperature estimation in annealers}
\author{Jack~Raymond\footnote{jack.raymond@physics.org}}
\author{Sheir~Yarkoni}
\author{Evgeny~Andriyash}
\affil{D-Wave Systems Inc., Burnaby, B.C, Canada}
\date{\today} 
\setcounter{Maxaffil}{0}
\renewcommand\Affilfont{\itshape\small}

\maketitle

\begin{abstract}
Sampling from a Boltzmann distribution is NP-hard and so requires heuristic approaches. Quantum annealing is one promising candidate. The failure of annealing dynamics to equilibrate on practical time scales is a well understood limitation, but does not always prevent a heuristically useful distribution from being generated. In this paper we evaluate several methods for determining a useful operational temperature range for annealers. We show that, even where distributions deviate from the Boltzmann distribution due to ergodicity breaking, these estimates can be useful. We introduce the concepts of local and global temperatures that are captured by different estimation methods. We argue that for practical application it often makes sense to analyze annealers that are subject to post-processing in order to isolate the macroscopic distribution deviations that are a practical barrier to their application.
\end{abstract}

\section{Introduction}
\label{sec:Introduction}
Boltzmann distributions are important in many areas of science, and sampling from these distributions is a major bottleneck in many interesting applications. The tasks of uniform generation, approximate counting, and inference (e.g., estimation of marginal probabilities), are often NP-hard~\cite{Sinclair198993,cooper1990computational,long2010restricted}. Heuristic samplers that sample approximately from a Boltzmann distributions are applied in practice to large scale problems (for example in machine learning~\cite{SalHinton07}).

One approach to heuristic sampling is to use an annealer. Whether thermal or quantum, an annealer generates independent samples by slowly transforming an easily prepared initial state into a random final state associated to a given objective function~\cite{Kadowaki:QA,Kirkpatrick:OSA}. In the case of the simulated thermal annealer (STA), an initial random sample is evolved through a schedule of decreasing temperature towards a specified terminal temperature~\cite{Kirkpatrick:OSA,Landau:GMC}. In quantum annealing the initial state is a ground state of some driver Hamiltonian (often a unifrom superposition of states). During the annealing process the state is evolved by slowly changing the Hamiltonian towars the the target Hamiltonian~\cite{Kadowaki:QA}. 

Although annealers have primarily been considered in the context of optimization, they can also be used as heuristic samplers of Boltzmann distributions. With sufficient resources, STA samples from a Boltzmann distribution~\cite{Kirkpatrick:OSA,Neal:PIMC}. However, the resources required per sample to achieve this are prohibatory in interesting applications, so that it is typically run as a heuristic without theoretical guarantees. Previous studies have also indicated that samples produced by the D-Wave quantum annealers may produce samples well described by finite temperature Boltzmann distributions~\cite{Bian:IM,Denil:TI,Dumoulin:CPI,PerdomoOrtiz:EET,Amin:QBM,Amin:SQS}.

In this paper we investigate several methods for determining how close sample distributions produced by annealers are to a family of Boltzmann distributions parameterized by inverse temperature $\beta$. These methods estimate the parameter $\beta$ best describing samples drawn from an annealer, and also provide measures of closeness. The annealers we evaluate are the latest-model D-Wave\footnote{D-Wave and D-Wave 2X are trademarks of D-Wave Systems Inc.} quantum annealer -- the D-Wave 2X (DW2X)~\cite{King:DDHT,King:BQA,Johnson2011Quantum,Denchev:WCV}, and an implementation of simulated thermal annealing on a single CPU. We consider two knobs for each annealer that modify the heuristic distributions generated: rescaling the STA terminal temperature/DW2X terminal energy scale, or changing annealing time (either in the DW2X or STA). Sample quality shown here does not reflect performance of optimally-tuned versions of these annealers, and are simply presented to compare various $\beta$ estimation techniques. 

We observe a significant difference between \textbf{local} (subspace) and \textbf{global} (full space) features of the annealer distributions. We find that even though samples are locally similar to a Boltzmann distribution, the global deviation can be large. This gives rise to a ``global warming'' effect: the fact that global distributional features indicate a higher temperature than local distributional features. We consider several estimators of inverse temperature and evaluate their efficacy. Some estimators are sensitive to details of the dynamics, and indicate a significant difference between the DW2X and STA. Other estimators are more sensitive to ergodicity breaking and macroscopic distribution features, where the DW2X and STA show a qualitatively similar behavior.

We treat our heuristic samplers as black-boxes and consider temperature estimation as the problem of determining the best fit amongst a single-parameter exponential family of models. This problem has a long history, and best practice is well established~\cite{Wainwright:GME,Geyer:MLE,Lehmann:TheoryofPointEstimation}. Inference of Ising models parameters under some systematic schemes is NP-hard~\cite{Bresler:HPE,Wainwright:GME}. However, heuristic approaches such as log-pseudo-likelihood are known to perform well in practice~\cite{Besag:PL}, and some schemes are provably convergent with reasonable resources~\cite{Montanari:CI,Bhattacharya:IIM}. Bhattacharya et al.~\cite{Bhattacharya:IIM} recently considered the log-pseudo-likelihood estimator for $\beta$ and found that estimation based on only a single sample is possible; their focus was primarily on the convergence properties of this estimator. Multi-parameter estimation (estimation of couplings and fields) is more commonly studied, and is pertinent to the class of Ising models we study, though beyond the scope of this paper. In this context, efficient methods of estimation for strongly interacting models include pseudo-likelihood and variational approaches~\cite{Albert:IIP,Aurell:IIP,Nguyen:MFT}.

Many recent papers have shown that physical quantum annealers approximate Boltzmann distributions~\cite{Bian:IM,Denil:TI,Dumoulin:CPI,PerdomoOrtiz:EET,Denil:TI,PerdomoOrtiz:DCPB,Amin:QBM,Amin:SQS}. In some of these approaches temperature estimators have been developed, and these estimators have been effectively applied in correcting the annealer parameterization to produce the desired distribution. A significant focus has been the impact of noise, or systematic specification errors, in D-Wave processors. Remedies have been proposed to allow more effective sampling, but scaling is either poor or unproven; in some methods only a restricted set of problem classes are appropriate. An extension to the temperature estimation method of Benedetti et al.~\cite{PerdomoOrtiz:EET} is discussed in supplementary materials, but we prefer the more standard estimators presented in the main text. Some work considering closeness to quantum Boltzmann distributions has appeared~\cite{Amin:SQS,Amin:QBM}. 

Our paper evaluates several standard methods, but differs from previous studies in that it uses insight specific to annealers in the analysis of deviations and development of temperature estimators. Noise sources and quantum features in physical quantum annealers are discussed only briefly. Some estimators we evaluate have a firm theoretical basis, such as maximum likelihood, but where this is lacking we will not focus on formal properties such as convergence, bias and variance. 

\begin{figure}[htb]
  \centering
  \includegraphics[width=0.6\textwidth]{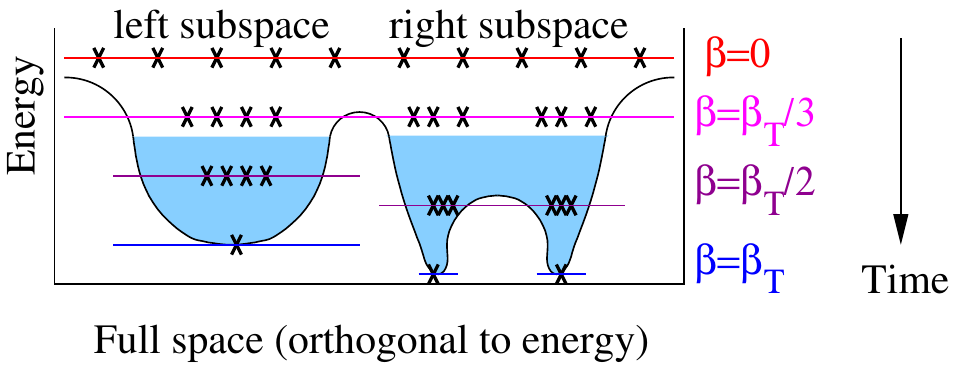}
  \caption{\label{fig:schematic} This is a schematic picture to illustrate ergodicity breaking in the STA. Proceeding through the anneal, samples evolve according to a schedule from some initial $\beta$ to the terminal value. In Boltzmann distributions there is an equivalence between mean energy and $\beta$~\cite{Wainwright:GME}. For illustration purposes we take all samples (x) to be concentrated about this mean energy, and show a qualitative distribution over the remainder of the high dimensional space. Initially ($\beta=0$) samples are uniformly distributed. For small $\beta$, samples equilibrate and can explore the entire space on short dynamical time scales. At some later time (in the schematic: $\beta_T/3$) the space may be partitioned into subspaces by energy barriers. At this point, samples can mix rapidly on the left subspace, or the right, but not between. For larger $\beta$, in the blue region, dynamics are too slow to allow mixing between the left and right subspaces (ergodicity is broken). The number of samples trapped in each valley is {\em approximately} controlled by the distribution at the earlier time ($\beta_T/3$) when mixing was still possible. At some later time again ($\beta_T/2$) mixing continues within each subspace. Due to the emergence of a second energy barrier, dynamics become slow on the right subspace (ergodicity is broken again on the right space). Finally, at $\beta_T$, the samples are distributed on low energy states, and if $\beta_T$ is large then all samples converge to their respective local minima. If ergodicity were not broken all samples would converge upon the global minimum. Note that, after ergodicity breaking, each subspace can have a distinct characteristic energy.}
\end{figure}

Qualitatively, the deviation of the STA distributions from the Boltzmann distribution for hard-to-sample Hamiltonians has been understood within physics and computer science since the idea of annealers was conceived~\cite{Kirkpatrick:OSA,Landau:GMC,Neal:PIMC}. 
As we modify the inverse temperature in the STA from its initial value to the terminal inverse temperature value ($\beta_T$), we move from a distribution where classical states are uniformly distributed to a distribution divided into disjoint subspaces of low energy (which can be identified qualitatively with modes of the probability distribution, or valleys in the free energy landscape). Under annealing dynamics, a state localized in one subspace cannot easily transition into another subspace that is separated by an energetic barrier once the inverse temperature becomes large (late in the annealing procedure) --- this is called \emph{ergodicity breaking}~\cite{Landau:GMC,Neal:PIMC}. In the case of the STA we gradually decrease the annealing temperature. Temperature is in one-to-one correspondence with expected energy in a Boltzmann distribution, and equilibrated samples are characterized by tight energy ranges. These samples are partitioned into subspaces by the energy barriers as temperature decreases, at which point the samples in each subspace will evolve independently, and be characterized by a local distribution. Rare fluctuations do allow samples to cross barriers, but are exponentially suppressed in the height of the energy barrier later in the anneal. Therefore, the distribution between subspaces will reflect the distribution at the point in the anneal where dynamics between the subspaces became slow, rather than the equilibrated distribution associated to the terminal model. This effect is called ``freeze-out''. We provide a schematic in Figure \ref{fig:schematic}. 

Similarly, in the case of quantum annealing, we proceed through a sequence of quantum models of decreasing transverse field (and increasing classical energy scale). With respect to the terminal diagonal Hamiltonian, the energy is again decreasing throughout the anneal, and some characteristic mean energy defines the sample set at intermediate stages. Energy barriers become impassable as the transverse field weakens and tunneling becomes slow, so that the process of ergodicity breaking is qualitatively similar~\cite{Kadowaki:QA,Amin:SQS}. Tunneling dynamics are affected by energy barriers in a different manner to thermal excitation dynamics, which is why there is some enthusiasm for quantum annealing; for some problems it may not suffer the same dynamical slowdown that is true of STA~\cite{Denchev:WCV}. 

For many problem classes, these points of ergodicity breaking become well defined in the large system size limit, and can often be directly associated to thermodynamic phase transitions~\cite{Landau:GMC,Mezard:IPC}. 

In this paper we consider heuristic sampling from classical Ising spin models. The state $x$ will consist of $N$ spins, defined on $\{-1,+1\}^N$. The Hamiltonian is
\begin{equation}
  H(x) = \sum_{ij} x_i J_{ij} x_j + \sum_i h_i x_i \label{eq:IsingHamiltonian},
\end{equation}
where $J$ and $h$ are unitless model parameters called couplings and fields respectively. The Boltzmann distribution corresponding to this Hamiltonian at inverse temperature $\beta$ is
\begin{equation}
  B_\beta(x) = \frac{1}{Z(\beta)} \exp(-\beta H(x)) \label{eq:PBoltzmann}\;,
\end{equation}
where $Z$ is the partition function. Throughout the paper we will use the standard, though improper, abbreviation in which $x$ can denote both the random variable $X$, and its realization. We study two problem classes compatible with the Chimera topology described in Section \ref{sec:models}, tailored to the D-Wave architecture~\cite{Ronnow:DTQS,Katzgraber:GC,King:BQA}. 

\subsection{Outline}
\label{sec:Outline}
In Section \ref{sec:tempEstimators} we introduce Kullback-Leibler divergence and mean square error on correlations as measures of closeness to the Boltzmann distribution, and from these develop standard estimators of inverse temperature.

In Section \ref{sec:LDF} we develop local self-consistent approximations for efficiently evaluating our inverse temperature estimators; we call these \emph{locally-consistent} inverse temperature estimators. We argue that in the context of annealers the approximation may determine an inverse temperature significantly different (typically larger) than that obtained by a full (computationally intensive) evaluation method. However, we will show that the estimate has meaning in that it captures local distribution features. Applying our approximation to maximum likelihood estimation, we recover in a special case the commonly used pseudo-log-likelihood approximation method. 

In Section \ref{sec:PP} we argue for evaluating post-processed distributions in place of raw distributions, for the purposes of removing superficial deviations in the heuristic distribution, and for determining the practical (as opposed to superficial) limitations of heuristic annealers. 

We then present experimental results relating to our objectives and estimators in Section \ref{sec:Experiments}. We conclude in Section \ref{sec:discussion}.

In supplementary materials we present various supporting results to complement the main text, and additional results and methods. In particular we introduce a new {\em multi-canonical} approximation for $\beta$ estimation inspired by Benedetti et al.~\cite{PerdomoOrtiz:EET}, we address issues related to practical usage of a DW2X, and we develop a method for calculation of KL-divergence. As part of our work we study the RAN1 and AC3 models of the main text, as well as two problem classes not tailored to the DW2X architecture~\cite{Douglass:QSAT,Kosko88bidirectionalassociative}. 

\subsection{Practical guidelines for using annealers in sampling applications}
Based on the results in this paper we offer the following advice for selecting and interpreting temperature estimation methods in the context of samples drawn from an annealer.
\begin{itemize}
\item It is important to define a suitable objective that is minimized by Boltzmann samples, and check that the objective is indeed small for the heuristic sampler for some temperature. It is not sufficient to find the best temperature, since there will always be a best temperature even for bad distributions. Comparisons of $\beta$ estimates between heuristics are not meaningful in the absence of this analysis. 
\item Robust evaluation of a heuristic sampler will often require input from an independent (exact or heuristic) method, such as statistical estimates against which to compare. Attempting to quantify error in a locally self-consistent manner could be misleading. 
\item The DW2X rescaling parameter/STA terminal temperature, total annealing time, and post-processing, should be tuned to the sampling objective. 
\item A temperature can be estimated accurately and efficiently by standard methods, such as the log-pseudo-likelihood method. If ergodicity breaking is a weak effect, the log-pseudo-likelihood estimator is sufficient.
\item It is valuable to consider several different types of estimator, since different estimators may be sensitive to different distribution features. Disagreement amongst estimators may reveal a pattern of ergodicity breaking, or imply a path to error mitigation. 
\item Efficient post-processing can move the distribution towards Boltzmann distribution by correcting local deviations. We are interested in the best pracitical heuristic, and so efficient post-processing should be applied.
\end{itemize}
We argue in this paper, in line with previous literature and experimental results, that the temperature that best describes an annealer distribution is expected to be a function not only of the annealer parameterization, but of the target Hamiltonian. There is no single parameter $\beta$ that is optimal for all Hamiltonians. Whilst this should be bourne in mind, closely related Hamiltonians (e.g. those of a given class, created by a slow learning procedure, or otherwise of comparable statistical properties) do yield comparable estimates for temperature, so that it may be efficient to estimate temperature properties on a small subset of the problems of interest and effectively generalize.

\section{Estimators for temperature}
\label{sec:tempEstimators}
\begin{figure}[htb]
 \centering
  \includegraphics[width=0.45\textwidth]{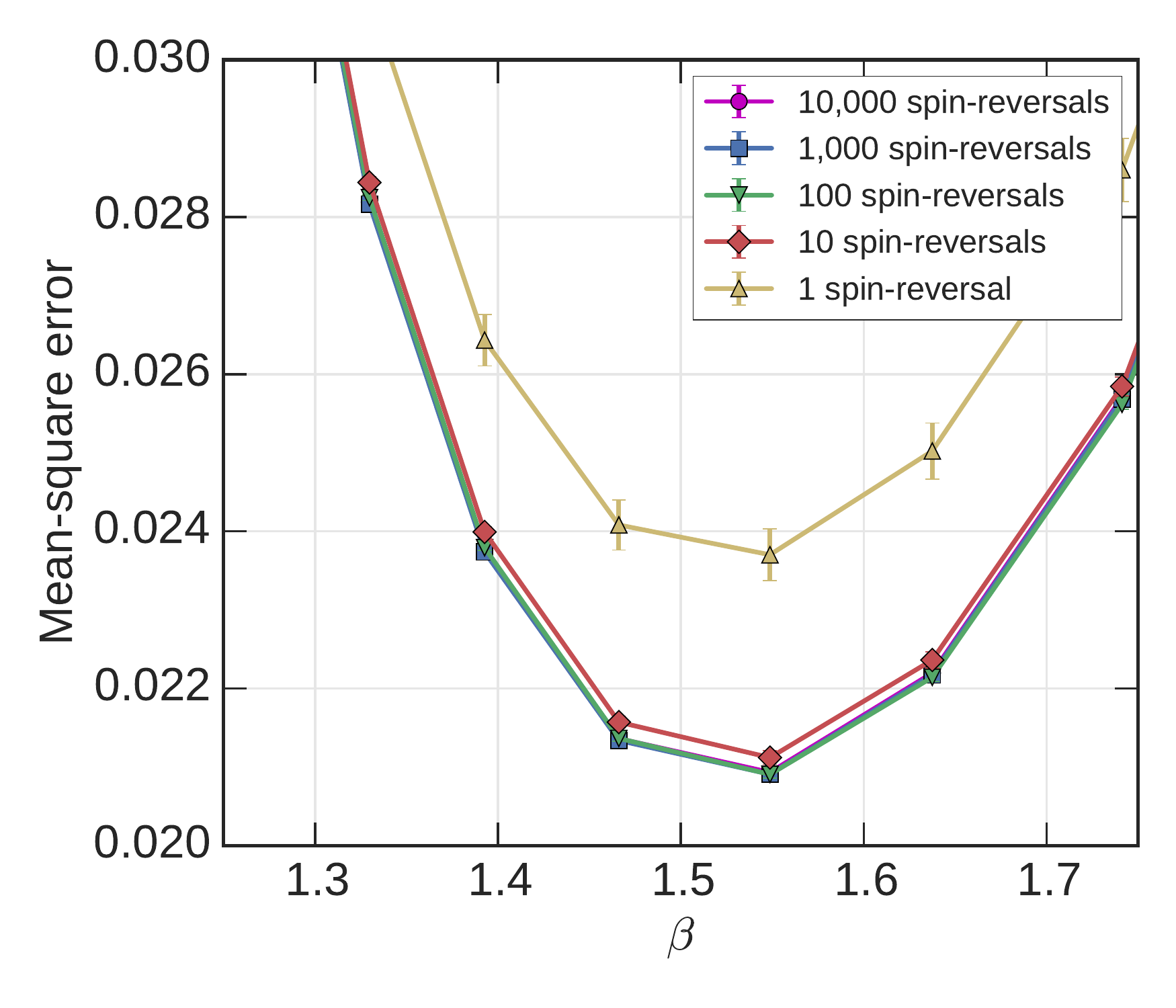}
  \includegraphics[width=0.45\textwidth]{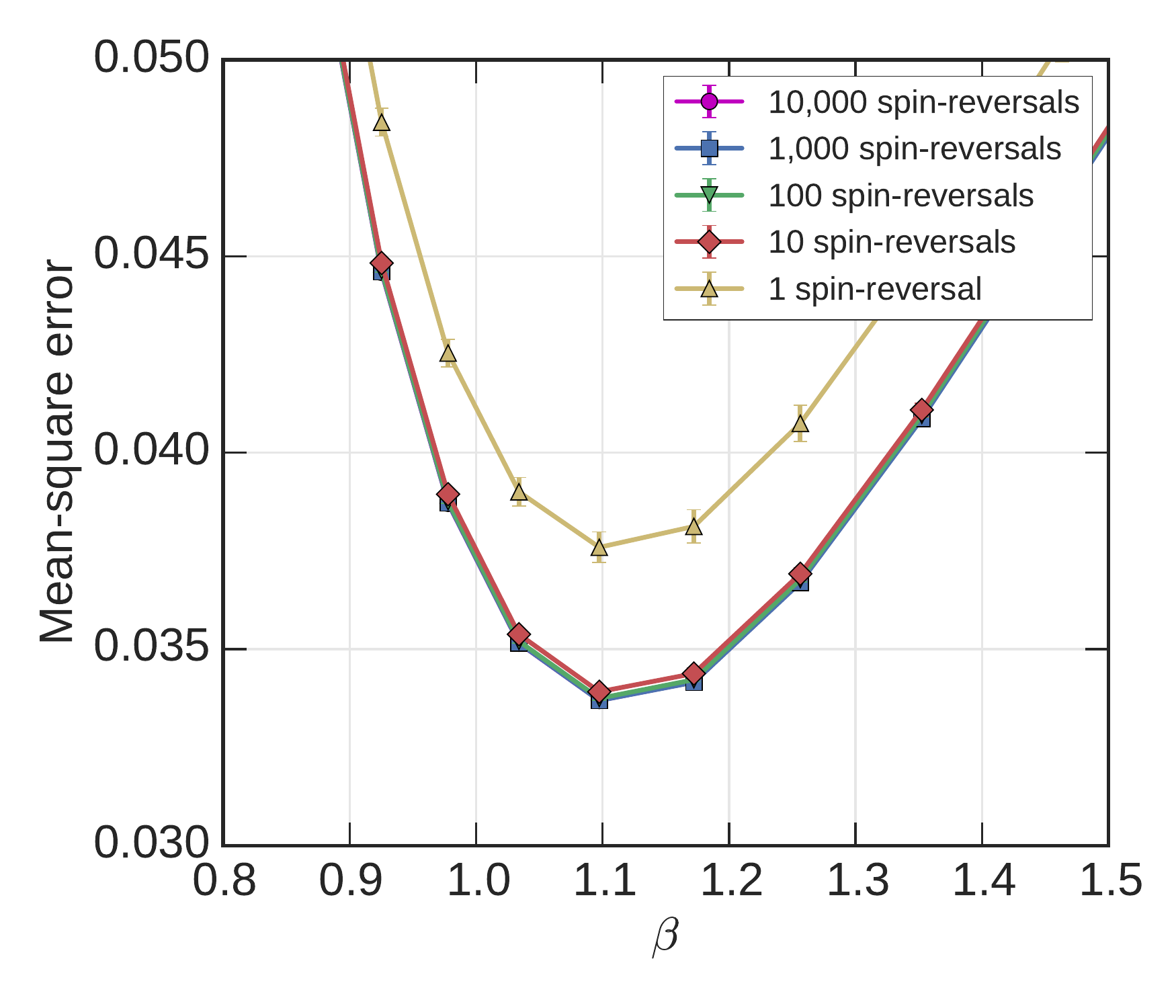}
  \caption{\label{fig:spinReversals} Spin reverals are a noise mitigation technique, described in supplementary materials. Choosing batches with more spin-reversals aids sampling quality bringing us closer to the paradigm of independent and identically distributed samples, but at the price of additional programming time. At C12 scale, spin reversals have a significant impact on MSE for both AC3 (left) and RAN1 (right). Given $M$ samples, the error achieved by a batch methods (m samples per spin-reversal with M/m spin reversals) is already close to that of the ideal scenario of one sample per spin reversal when $m=10$. The signal is noisy when few spin-reveral transformations are used, and is only statistically significant after averaging over many distributions ($100$ in this figure). We establish the mean estimate, and its standard error by bootstrapping of a sample set of $10000$ spin-reverals sampling each time $1000$ samples.}
\end{figure}

We assume that annealers generate independent and identically distributed samples, according to a distribution $P_A$. For the STA this is reasonable given powerful pseudo-random number generators. For the DW2X, correlated noise sources (discussed in Section \ref{sec:DW2X}) means this is an approximation that is more difficult to analyze. In Figure \ref{fig:spinReversals} we show evidence that these weak correlations in time do not strongly affect our results and conclusions. The experimental structure is demoted to supplementary materials. 

We are interested in comparing these heuristic distributions to a family of Boltzmann distributions (\ref{eq:PBoltzmann}) parameterized by inverse temperature $\beta$. Amongst such models we wish to find the best fit, and measure its goodness. We will consider the best temperature to be that which minimizes some objective function. Since the distribution $P_A$ is a heuristic distribution, and not Boltzmann, this temperature may vary between objectives. Given an objective that is minimized at some unique inverse temperature, we then need an estimator for this temperature working on the basis of finite sample sets. An effective estimator should be consistent, with low bias and variance~\cite{Geyer:MLE,Lehmann:TheoryofPointEstimation,Bhattacharya:IIM,Shirts:SOA,Montanari:CI}. The estimators we study will be consistent, and in some cases optimal with respect to variance and bias (e.g. the Maximum Likelihood estimators~\cite{Geyer:MLE,Lehmann:TheoryofPointEstimation}).

Either to evaluate the objective or to estimate temperature (i.e. minimize the objective) note that we must evaluate some statistics from the Boltzmann distribution, e.g., the mean energy, an energy gap, or marginal distribution. Inference for any of these quantities is NP-hard in the model classes we study. It is often in practice easier to evaluate the energy, and perhaps $\log(Z)$, than marginal statistics, but estimation of all these quantities is slow in the worst case. 
For purposes of the models and temperatures explored we are able to accurately estimate the mean energy, $\log(Z)$, or marginal expectations under the Boltzmann distribution by either dynamic programming or parallel tempering~\cite{Wainwright:GME,Hukushima:EMC,Selby:sampling}. With these values in hand we can efficiently evaluate our temperature estimators, and in most cases the objective (an exception, Kullback-Leibler divergence, is discussed in supplementary materials). However, a scalable estimator requires us to find effective approximation methods for these quantities or to define different estimators, and is the subject of Section \ref{sec:LDF}.

\subsection{Maximum likelihood (Minimum Kullback-Leibler divergence)}
\label{sec:MLE}
When comparing distributions, a natural objective function to minimize is the Kullback-Leibler (KL) divergence between the sampled distribution (from the annealer) $P_A$ and the corresponding Boltzmann distribution $B_\beta$, as follows:
\begin{equation}
  \mathrm{D}_{KL}[P_A,B_\beta]= \sum_x P_A(x) \log\left(\frac{P_A(x)}{B_\beta(x)} \right) \label{eq:KLD}.
\end{equation}
The Kullback-Leibler divergence is an important information-theoretic quantity, which places various limitations on the efficacy of $P_A$ for modeling $B_\beta$, and vice-versa~\cite{Wainwright:GME}\footnote{The reverse form of the KL-divergence, or its symmetrized form, are also interesting. We choose this form as it allows for evaluation in the limit $P(x)\rightarrow 0$, amongst other technical factors. This is discussed further in supplementary materials.}.

$P_A$ has no $\beta$ dependence, so that at the minimum of this function with respect to $\beta$, we obtain an energy matching criterion $\mathrm{EM}(\beta)=0$, where
\begin{equation}
  \mathrm{EM}(\beta) = \sum_x P_A(x) H(x) - \sum_x B_\beta(x) H(x) \label{eq:EM}.
\end{equation}
The energy matching criterion yields the maximum likelihood estimator for $\beta$ -- the likelihood that the annealed samples were drawn from a Boltzmann distribution. Maximum likelihood is perhaps the most well established of procedures for estimating model parameters from data -- in this case the data being the samples drawn from $P_A$~\cite{Lehmann:TheoryofPointEstimation,Geyer:MLE,Wainwright:GME}. Note that the Boltzmann distribution is an exponential model, and so it is natural to define the estimator in terms of expected energy, which is the sufficient statistic associated to the parameter $\beta$~\cite{Wainwright:GME}. 

\subsection{Minimum mean square error on correlations}
\label{sec:MSE}
In the context of machine learning, an important potential application of annealers, the important feature of samples may be the quality of some statistics that are derived from them. In particular, a machine learning process may require accurate estimation of single variable expectations, and expectations for products of variables (correlations). For this reason we consider an alternative objective, the mean square error (MSE) on correlations:
\begin{equation}
  \mathrm{MSE}[P_A,P_\beta]= \frac{1}{M}\sum_{ij : J_{ij}\neq 0}\left( \sum_x [P_A(x)-P_\beta(x)] x_i x_j \right)^2 \label{eq:MSE}\;,
\end{equation}
$M$ is number of non-zero couplings.
We consider specifically the mean error on correlations (excluding errors on single variable expectations) since the models we study experimentally are all zero-field problems so that $E[x_i]=0$ for all $\beta$ by symmetry. 
Unlike the KL-divergence, MSE is not a convex function of $\beta$ in general, although intuitively this might be expected for many problem classes and reasonable heuristics. A derivative of (\ref{eq:MSE}) with respect to $\beta$ will yield a criterion for local optimality. This is a complicated expression dependent on many statistics, but is straightforward to approximate numerically in our examples. 

We will find that in application to annealers the minimum for this second objective (\ref{eq:MSE}) can disagree with the maximum likelihood (minimum KL-divergence) estimator (\ref{eq:EM}), typically being more sensitive to ergodicity breaking in our experiments. Note that, once ergodicity breaking has occurred, the mean energy can be improved as samples settle towards their respective local minima. The maximum likelihood value can be larger than that implied at the point of ergodicity breaking. By contrast, the distribution between the now disconnected subspaces that determines the correlations cannot be much improved as samples settle towards their local minima. Therefore the minimum MSE estimator will typically indicate a smaller value for $\beta$, better inline with the point of ergodicity breaking. We stress that MSE is not in any sense a special objective function in this regard; many variations are possible and should be chosen in an application orientated manner.

\section{Local approximations for evaluating objectives and estimators}
\label{sec:LDF}

The problem with the objectives and estimators outlined in Section \ref{sec:tempEstimators} is that their use requires inference with respect to the Boltzmann distribution that is independent of the heuristic annealer, which is NP-hard to perform: estimation of either the expected energy, or correlations.

In this section we show how, beginning from the annealed distribution, we can build a reasonable approximation to the Boltzmann distribution and thereby evaluate the estimator self-consistently. The estimators are motivated as approximations to those of Section \ref{sec:tempEstimators}. However, we will show that even in cases where the approximation is poor, the estimator can still reveal useful information about the distribution.

\subsection{Statistics of the heuristic distribution, and of the Boltzmann distribution, by local self-consistency}
\label{sec:SCM}

Our estimators and objectives require us to evaluate statistics of the annealed distribution $P_A$. Estimates of mean energy or correlations based on $P_A$ can be obtained by evaluating those statistics from the sample set $\mathcal{S}=\{x\}$, or equivalently, evaluating their corresponding expressions using the plug-in estimator to $P_A$:
\begin{equation}
  {\hat P}_A(x) = \frac{1}{|\mathcal{S}|} \sum_{x' \in \mathcal{S}} \delta_{x,x'} \label{eq:PlugIn}\;.
\end{equation}
The quality of estimates depends on variance and sample size. In experiments we typically use sample sets of size $10^4$ that are sufficient for temperature estimation and evaluation of the objectives.
The evaluation of the KL-divergence is one exception: our approximation (\ref{eq:PlugIn}) is known to fail when applied to the entropy term $-\sum_x P_A(x) \log P_A(x)$~\cite{Paninski:EE,Grassberger:EE}. In the supplementary materials we discuss why evaluation is problematic, and propose a mitigation strategy.

We must also evaluate energy, correlations and $\log(Z)$ under the Boltzmann distribution, which is NP-hard. However, under the assumption that $P_A(x)$ is close to the Boltzmann distribution we may make a {\em locally-consistent} approximation. The approximation to $B_\beta(x)$ is
\begin{equation}
  {\hat B}_\beta(x) \propto \sum_{x'} {\hat P}_A(x') W_\beta(x|x') \label{eq:BlocalApprox},
\end{equation}
where $W_\beta$ is a $\beta$-dependent kernel. It efficiently maps any state into a new state, with the property that the distribution is unchanged if it is a Boltzmann distribution
\begin{equation}
  B_\beta(x) \propto \sum_{x'} B_\beta(x')W_\beta(x|x') \label{eq:consistentkernel}.
\end{equation}

The transition kernels used in Markov chain Monte Carlo (MCMC) methods, either singly or iteratively, are suitable candidates for $W_\beta$~\cite{Landau:GMC,Mezard:IPC}. The blocked Gibbs MCMC method is described in Section \ref{sec:BlockedGibbs}. The simplest example of $W_\beta$ is conditional resampling of a single variable, which is an element in the blocked Gibbs sampling procedure. All variables except $i$ are unchanged, and $i$ is resampled according to the conditional Boltzmann distribution $B_\beta(x_i|x \setminus x_i)$ given the neighboring values. We label this kernel by $(i)$, indicating the updated variable
\begin{equation}
  W^{(i)}_\beta(x|x') = B_\beta(x_i | x' (\neq x_i') ) \prod_{j (\neq i)} \delta_{x_j,x_j'} \label{eq:W1var}\;.
\end{equation}
Applying the approximation (\ref{eq:BlocalApprox}) in combination with the kernel (\ref{eq:W1var}) to maximum likelihood estimation we obtain an energy matching criterion for $(i)$. Each kernel $i$ defines an energy matching criterion and an estimate for $\beta$, but it is not possible to simultaneously satisfy the criteria for all $i$. We can make a composite energy matching criterion by weighting each of the criteria equally: taking an average over $\mathrm{EM}(\beta)$ for each $i$. In this case we recover the maximum log-pseudo-likelihood (MLPL) estimator~\cite{Besag:PL,Bhattacharya:IIM}. The MLPL estimator is normally derived and motivated slightly differently. An alternative way to combine the kernels $\{W^{(i)}\}$ is to define a composite kernel as a sum of the individual kernels. We prefer the MLPL estimator in this paper due to its prevalance in the literature and well established statistical properties. We discuss this further in supplementary materials, where alternative locally self-consistent estimators are also examined.

In the case of the Hamiltonian (\ref{eq:IsingHamiltonian}) the MLPL estimate is the solution to $\mathrm{EM}(\beta)=0$, where
\begin{equation}
  \mathrm{EM}(\beta) = \sum_{x \in \mathcal{S}} \sum_i x_i \zeta_i(x) \frac{ \exp(2 \beta x_i \zeta_i(x) )}{1 + \exp(2\beta x_i \zeta_i(x) )} \label{eq:MLPL}\;,
\end{equation}
where $\zeta_i(x) = [h_i + \sum_j (J_{ij}+J_{ji})x_j]$ is the effective field. Note that $-2 x_i \zeta_i$ is the energy change of flipping the state of spin $i$. Provided there exists at least one $\zeta_i(x)>0$, and one value $\zeta_i(x)<0$ (at least one local excitation in some sample), then this equation has a unique finite solution which can be found, for example, by a bisection search method.

For our purposes, the MLPL estimator is a special case of a more general {\em locally consistent} estimator. We choose a kernel, approximate $B_\beta$ (\ref{eq:BlocalApprox}), and then evaluate the energy matching criterion (\ref{eq:EM}). The locally consistent approach can also be applied straightforwardly to the MSE, and minimum MSE estimator, of Section \ref{sec:MSE}. In the main text the only locally consistent estimator for which we present results is the MLPL estimator. In the supplementary materials we perform an experiment to demonstrate how the strength of the kernel impacts temperature estimation. 

Consider the following interpretation for the role of the kernel: We take every sample that the annealer produces, and conditionally resample according to $W_\beta$. We then take this new set of samples as an approximation to Boltzmann samples drawn according to $B_\beta$. Since $W_\beta$ obeys detailed balance, it necessarily brings the distribution towards the Boltzmann distribution. Consider again Figure \ref{fig:schematic}, and note that resampling single spins, or doing some other efficient conditional resampling procedure (i.e. some short-run MCMC procedure) does not lead to a significant macroscopic redistribution of the samples, except in the high-temperature regime where fast dynamical exploration of the space is possible. Thus, the approximation (\ref{eq:BlocalApprox}) will typically inherit the macroscopic bias of the sampling distribution through ${\hat P}_A$, but correct local biases. The locally consistent estimator is therefore effective in capturing any {\em local} deviation in the distribution $P_A$ not representative of $B_\beta$.

\subsection{Towards global distribution features}
\label{sec:PP}
Objective functions such as maximum likelihood and minimum mean square error on marginals are influenced by a combination of local and global distribution features. We have already seen that locally consistent estimators, such as MLPL, can assign a meaningful temperature for local deviations from the Boltzmann distribution. However, these may fail to capture macroscopic features. It would be useful to have an objective, or estimator for temperature, that reflects only the macroscopic distribution features. One way to do this is to manipulate $P_A$ so that the local distributional features are removed.

We have also not considered so far the practical application of annealers as heuristic samplers. By our definition, for an annealer to be useful it must do well on the appropriate objective, and be fast. However, these two aims are typically in tension. A method that allows one to trade off these two goals is post-processing. In post-processing we take individual samples, or the set of all samples, and apply some additional procedures to generate an improved set of samples. This requires additional resources and can be heuristic, or employed in a manner guaranteed to improve the objective. 

Those distribution features that can be manipulated by post-processing will be called local. Local, since it is assumed that efficient post-processing will not be so powerful as to manipulate the macroscopic distribution in interesting cases. Amongst the easiest local feature to correct in annealers is local relaxation: the tendency of states to decrease in energy towards their local minima at the end of the anneal as illustrated in Figure \ref{fig:schematic}.

Post-processing has two uses considered in this paper: to extract macroscopic distribution features (by discounting local distortions), and to improve the heuristic distributions. A post-processed distribution can be represented as
\begin{equation}
  P_{\beta,A}(x) = \sum_{x'} W_\beta(x|x')P_A(x') \label{eq:Ppp}\;,
\end{equation}
where $W_\beta$ is again a kernel. 

With post-processing, we now have three distributions of interest: $P_{A}(x)$, $P_{\beta,A}(x)$ and $B_\beta(x)$. Until this section we were interested exclusively in the closeness of $P_{A}(x)$ and $B_{\beta}(x)$, and under the assumption that $B_\beta\approx P_{\beta,A}(x)$ we developed efficient approximations to maximum likelihood (or minimum MSE) estimation in Section \ref{sec:SCM}. We can now present a different interpretation. If we are interested in local deviations in the distribution, we should compare $P_{A}(x)$, $P_{\beta,A}(x)$; whereas if we are interested in global deviations we should compare $P_{\beta,A}(x)$ and $B_\beta(x)$. For practical purposes only the latter comparison makes sense, since we can efficiently correct local errors. However, the former is important in understanding why annealers fail, and how to correct their distributions. The locally self-consistent estimators (such as MLPL) minimize a divergence between $P_{A}(x)$ and $P_{\beta,A}(x)$, and so should be interpreted as local approximations (yielding a local temperature estimate). In the case that $P_A(x)=B_\beta(x)$, then the distinction between global and local is no longer relevant, and this local approximation is a consistent and low variance estimator for the unique $\beta$ describing the distribution.

Implicit in our definition (\ref{eq:Ppp}) is a restriction to ``do no harm'' post-processing, we post-process at the $\beta$ that defines the Boltzmann distribution of interest (or that to which we wish to compare). The criterion (\ref{eq:consistentkernel}) does no harm since it is guaranteed to move any distribution towards a Boltzmann distribution in some sense, and never away from it. In the case that $W_\beta$ involves only conditional resampling according to $B_\beta$ (rule (\ref{eq:W1var}) is one such case) it is straightforward to show that the $\mathrm{D}_{KL}[P_{\beta,A},B_\beta] \leq \mathrm{D}_{KL}[P_A,B_\beta]$. It is reasonable to expect, though not guaranteed, that other objectives will improve under do no harm kernels. Heuristic approaches without such guarantees may sometimes do better in practice, but carry risks. 

For purposes of isolating macroscopic features of the distribution it is ideal to apply enough post-processing to remove the local distortions; but leave the macroscopic features intact. This is a balancing act that strictly exists only as a concept, since the distinction between local and global is blurred except perhaps in the large system size limit, and there will typically be several relevant scales not just two. In experiments we present results for post-processing consisting of one sweep of blocked Gibbs sampling (described in Section \ref{sec:BlockedGibbs}), a weak form of post-processing.
For purposes of improving the heuristic, $W_\beta$ should be chosen powerful enough that the time-per-sample is not significantly impacted. One sweep of blocked Gibbs sampling meets the criterion of being a small overhead in time per-sample for the DW2X and STA under the operation conditions we examine, so we can infer something of the power of post-processing to efficiently correcting annealer non-idealities.

If the heuristic distribution is a function of $\beta$ (\ref{eq:Ppp}), we must take into consideration the dependence of $P_{\beta,A}$ on $\beta$ in objective minimization. KL-divergence minimization becomes distinct from maximum likelihood in the case that samples are a function of $\beta$, and the energy matching criterion (\ref{eq:EM}) is modified in the former case. This point is further discussed in supplementary materials. 

In Section \ref{sec:LDF} we developed locally self-consistent estimators. We emphasize that with post-processing, these estimators can be made redundant, unless the post-processing method kernel (\ref{eq:Ppp}) is significantly different from the kernel used in the local self-consistency trick (\ref{eq:BlocalApprox}). The self-consistency trick (\ref{eq:Ppp}) uses information about how samples are redistributed under the kernel to determine $\beta$ - if this kernel matches the post-processing kernel, then it detects the effect of post-processing and very little else. Therefore, in the evaluation of post-processed distributions care should be taken in applying and interpreting self-consistent approximations to $\beta$.

\section{Experimental results}
\label{sec:Experiments}

In Section \ref{sec:models} we present the two models we will study in the main text. We then describe in Section \ref{sec:BlockedGibbs} a simple Markov Chain Monte Carlo procedure called blocked Gibbs sampling, and how we use this to create the STA. The blocked Gibbs method is also applied in our post-processing experiments. We then describe our usage of the DW2X as a sampler. Experimental results demonstrating the methods of Sections \ref{sec:tempEstimators} and \ref{sec:PP} are subsequently presented.



\subsection{RAN1 and AC3 models}
\label{sec:models}
\begin{figure}[htb]
 \centering
  \includegraphics[width=0.45\textwidth]{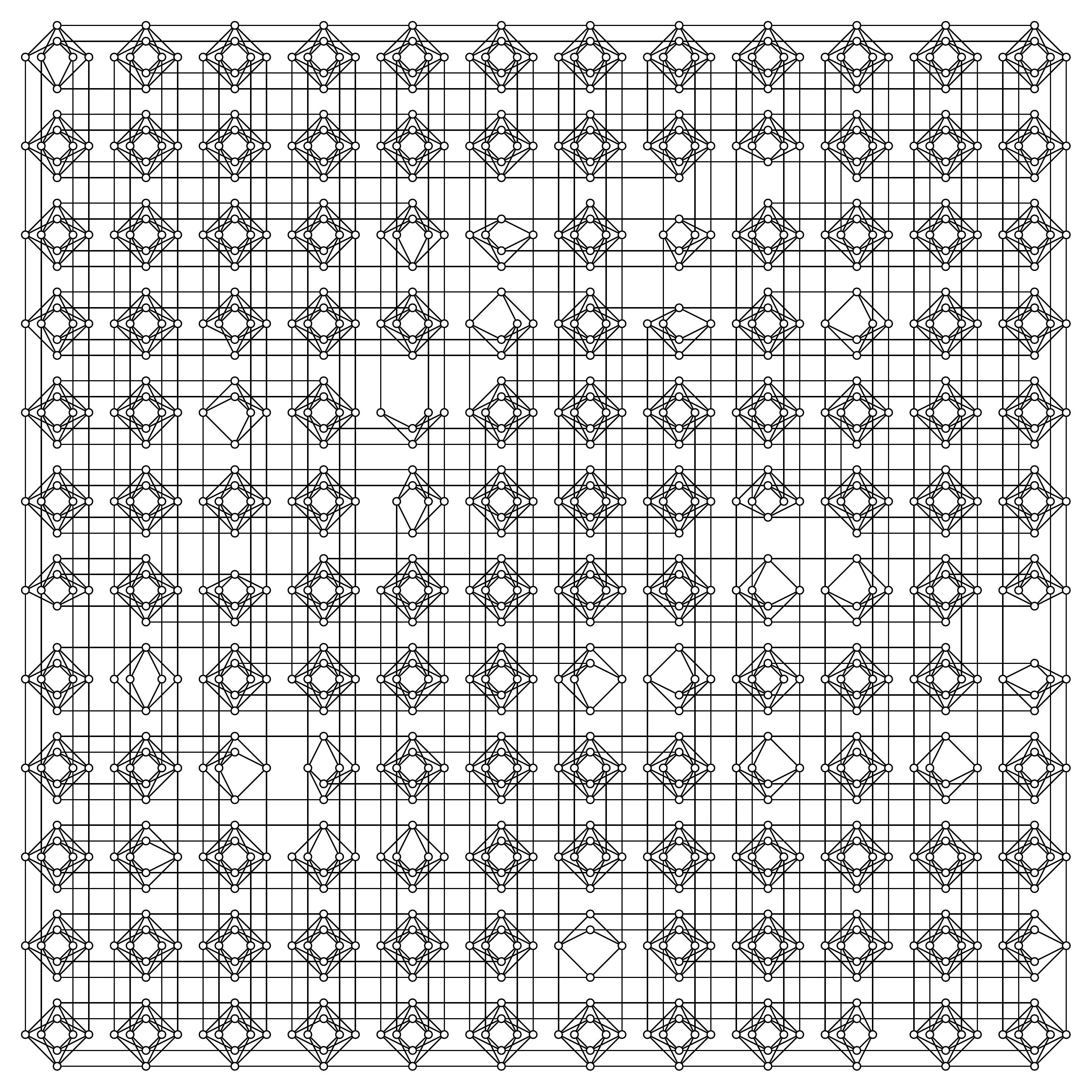}
  \caption{\label{fig:HW_Graph} Working graph of the DW2X used; the topology is called Chimera.}
\end{figure}

In this main section we consider only two models, RAN1 and AC3, which are spin-glass models compatible with the DW2X topology~\cite{Bunyk:AC}.
This topology is described by a {\em Chimera} graph shown in Figure \ref{fig:HW_Graph}; each variable is a circle with an associated programmable field $h_i$, and each edge is associated to a programmable coupling $J_{ij}$. Qubits are arranged in unit cells, each a $K_{4,4}$ bipartite graph of $8$ qubits. Due to manufacturing errors, some qubits and couplings are defective and cannot be programmed.

The RAN1 and AC3 problems are defined on a Chimera graph with $1100$ qubits across a $12$x$12$ cell grid (C12). In experiments we consider models that exploit all available couplings and qubits (C12), as well as models that use only $127$ qubits on a $4$x$4$ cell subgrid (C4), and $32$ qubits on a $2$x$2$ cells subgrid (C2). 

RAN1 is a simple spin-glass model without fields ($h_i=0$), and with independent and identically distributed couplings uniform on $J_{ij}=\pm 1$~\cite{King:BQA}. 
Recent work has indicated that for some algorithms RAN1 may be a relatively easy problem in which to discover optima~\cite{Ronnow:DTQS}, and that asymptotically there is no finite temperature spin-glass phase transition~\cite{Katzgraber:GC}, making it questionable as a benchmark. However, the problem class demonstrates many interesting phenomena at intermediate scales and has become a well understood benchmark for experimental analysis.

The AC3 model is a simple variation of the RAN1 class. Again we have no fields, but the intra-cell couplings' values (those between variables in the same cell) are sampled uniformly at random from $J_{ij}=\pm \sfrac{1}{3}$, and inter-cell couplings set to $J_{ij}=-1$~\cite{King:BQA}\footnote{We could equivalently assign the couplings between cells to $\pm 1$ at random, due to a simple symmetry the problem is not meaningfully changed.}. By making couplings relatively stronger between cells, longer range interactions are induced through sequences of strongly correlated vertical, or horizontally, qubits. If we can consider the inter-cell couplings to dominate energetically, then the low energy solution space becomes compatable with that of a bipartite Sherrington Kirkpatrick model~\cite{Venturelli:SK}. We find that the AC3 problem is an interesting departure from RAN1 since the solution space is less dependent on local interactions, and also because we modify the precision of couplings (from an alphabet of $\pm 1$, to an alphabet of $\pm \sfrac{1}{3}, -1$).

\subsection{Blocked Gibbs sampling, and simulated thermal annealing}
\label{sec:BlockedGibbs}
Blocked Gibbs is a standard Markov chain Monte Carlo procedure closely related to the Metropolis algorithm procedure~\cite{Landau:GMC,CP:ContrastiveDivergenceLearning}. It is the basis for the STA in this paper, and also the post-processing results.

First note that because the Chimera graph is bipartite, it is 2-colorable. Given a coloring, the variables in a set of a given color are conditionally independent given the variables of the complementary set. Thus it is possibility to simultaneously resample all states in one set, each according to the probability 
\begin{equation}
  P_\beta(x_i| x \setminus x_i) = \frac{\exp(- \beta \zeta_i(x) x_i)}{2\cosh(\beta \zeta_i(x))}\;,
\end{equation}
where $\zeta(x)=(J+J^T)x+h$. We proceed through the colors in a fixed order, for each color resampling all variables. An update of all variables is called a sweep. This procedure can be iterated at a fixed temperature, and the distribution of samples is guaranteed to approach the Boltzmann distribution parameterized by $\beta$ over sufficiently many sweeps. A graph coloring needn't be optimal, and can always be found (given sufficiently many colors), so that this algorithm generalizes in an obvious manner to non-bipartite graphs.

The blocked Gibbs sampling procedure at large $\beta$ is not very efficient in sampling for multi-modal distributions, since samples are immediately trapped by the nearest modes (which may be of high energy), and escape only over a long timescale. A more effective strategy for multi-modal problems is blocked Gibbs annealing, in which $\beta$ is slowly increased towards some terminal value $\beta_T$ according to a schedule (a schedule assigns one temperature to each sweep of blocked Gibbs). In this paper we consider an annealing schedule that is a linear interpolation between $0$ and $\beta_T$. Given the restriction to a linear schedule the STA has two parameters: the total anneal time, and the terminal inverse temperature $\beta_T$. The setting of these parameters is discussed in Section \ref{sec:DefaultParameterization}. 

\subsection{Quantum annealing with the D-Wave 2X}
\label{sec:DW2X}
\begin{figure}[htb]
  \centering
    \includegraphics[width=0.6\textwidth]{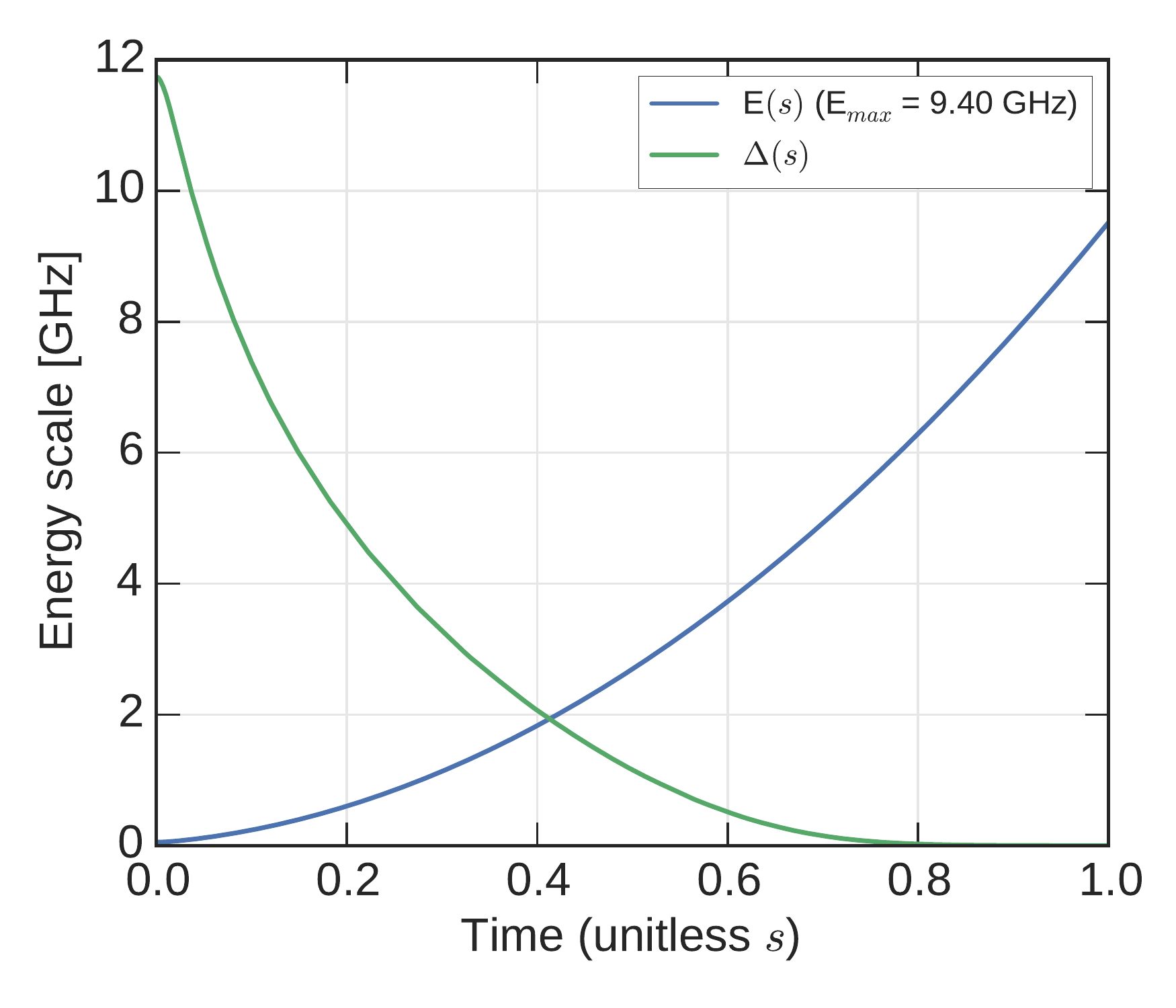}
    \caption{\label{fig:schedule} The DW2X operational energy scales during the anneal, the Hamiltonian is (\ref{eq:Hoperator}).}
\end{figure}

In the case of the DW2X, annealing is controlled by a time-dependent transverse field $\Delta(t)$ and an energy scale $E(t)$. These quantities are shown in Figure \ref{fig:schedule} for the DW2X used in this paper. The physical temperature ($T$) of the system varies with time and load on the device and is difficult to estimate. We have experimentally observed a physical temperature which is $22.9$ in the median, with quartiles of $22.0$~mK and $25.6$~mK, over the data collection period. We did not analyze the time scales associated to temperature fluctuations in depth, but much of the variation occurs on long time scales, so that in a single experiment we typically found a tighter range of temperatures applied.
The unitless Hamiltonian operator in effect during the anneal is
\begin{equation}
  {\hat H}(s) = \frac{\mathrm{h}}{k_B T}\left(\frac{\Delta(s)}{2} \sum_i {\hat \sigma}^{(i)}_x + \frac{E(s) r}{2} \left[\sum_{ij} \frac{J_{ij}}{\max(\{|J_{ij}|,|h_i|\})} {\hat \sigma}^{(i)}_z {\hat \sigma}^{(j)}_z + \sum_i \frac{h_i}{\max(\{|J_{ij}|,|h_i|\})} {\hat\sigma}^{(i)}_z\right]\right) \label{eq:Hoperator}
\end{equation}
where $s=t/t_{max}$ is the rescaled time, the coefficient $\mathrm{h}$ is Planck's constant, and ${\hat \sigma}$ are the Pauli matrices. The Hamiltonian parameters can be considered rescaled to maximum value $1$ (the function of the denominator $\max(\cdot)$). $r$ and $t_{max}$ are the rescaling parameter and the anneal time parameter, respectively. Throughout this paper, we adopt the convention that the Hamiltonian for the problem is fixed, and treat $r$ as a parameter of the heuristic (DW2X). In current operation of the DW2X, this manipulation is achieved in practice by turning off autoscale, and manually rescaling the values of $\{J_{ij},h_i\}$ submitted. We find the convention of modifying the quantum Hamiltonian (\ref{eq:Hoperator}) to be more intuitive than considering modification of the classical Hamiltonian that is submitted: when we rescale downwards we are implying the use of smaller energy scales $E(t)$ in the annealer, and we anticipate that $\beta$ is reduced.

\subsection{Parameterization of the DW2X and STA in experiments}
\label{sec:DefaultParameterization}
Our criteria for selection of the default rescaling parameter $r$ and the anneal time is minimization of the ``Time to Solution'' (TTS)~\cite{King:BQA,Ronnow:DTQS,TroyerSTA}. TTS is the expected time required to see a ground state for the first time. For the DW2X, it is optimal to choose a minimal programmable anneal time ($20\mu s$), and a maximum programmable energy scale $r=1$. A second (and not uncorrelated) reason to use these parameters is that they are the default operation mode of the DW2X. Since our objective is not optimization, TTS is not the optimal way to use the DW2X, but represents a standard choice that allows for phenomena to be explored.

The STA parameters are chosen in simple manner in part according to TTS, and in part for convenient comparison with the DW2X. We choose the STA parameter $\beta_T$ so that the distribution of local excitations in the DW2X and STA are comparable in the median case of 100 randomly generated C12 problems: by equating the {\em local} properties' differences between the annealers, we reflect more interesting {\em global} features. To equate local properties $\beta_T$ is chosen equal to the MLPL estimate of $\beta$ at full scale (C12) for the DW2X (see Figure \ref{fig:rescalingDW2X}). This implies $\beta_T=3.54$ for RAN1, and $\beta_T=4.82$ for AC3. In the case of RAN1 our choice $\beta_T=3.54$ is not too dissimilar to the value ($\beta=3$) that had been previously used for optimization applications~\cite{King:BQA, Ronnow:DTQS, TroyerSTA}.

We choose two sets for the number of sweeps of the STA, and show both in most figures.
The first set is chosen to minimize TTS in the median instance. For both RAN1 and AC3 we find an approximately linear trend in the width the Chimera graph, so that $12000$, $4000$ and $2000$ sweeps were close to optimal for C12, C4 and C2 problems respectively.

The second set was chosen to match the time per sample of the DW2X. A recent efficient implementation of the Metropolis algorithm for RAN1 problems achieves a rate of $6.65$ spin flips per nano-second~\cite{TroyerSTA}. In C$12$ problems we have $1100$ active qubits, and so $20\mu s$ would allow for $\frac{20000~ns }{1100~\mathrm{spin~flips}} * 6.65~\mathrm{spin~flips}/ns\approx 120$ sweeps (updates of all variables). For C4 and C2 we rescale linearly for simplicity ($40$ and $20$ respectively).

In experiments we evalute the STA and DW2X on the basis of sample batches, each batch consisting of $10^4$ samples.
We approximate the samples as independent and identically distributed. in the case of the DW2X this is an approximation and different programming procedures can impact quality of results. Our DW2X batches are generated by collecting $10^4$ samples split across $10$ programming cycles. This is a standard collection procedure that trades off quality of samples against timing considerations -- including annealing time, programming time and read-out time. The programming cycles exploit spin-reversals, a noise mitigating technique that strongly suppresses correlations between programming cycles~\cite{Boixu:Gauge}. The effect of this batch structure are presented in Figure \ref{fig:spinReversals}.

\subsection{Maximum likelihood and maximum log-pseudo-likelihood estimation}
\label{sec:MLandMLPL}

In this section we consider the DW2X and STA without post-processing. The maximum log-pseudo-likelihood (MLPL) estimate can be interpreted as a form of locally self-consistent approximation to maximum likelihood, as described in \ref{sec:SCM}, and here we compare it to the more computationally intensive maximum likelihood (ML) estimate of Section \ref{sec:MLE}. MLPL and maximum likelihood estimators indicate different values of $\beta$,  reflecting the failure of the locally self-consistent approximation (\ref{eq:consistentkernel}). MLPL captures a local temperature, consistent with the range over which the kernel (\ref{eq:W1var}) redistributes the sample. We consider the estimates with variation of the DW2X rescaling parameter $r$ and the STA terminal temperature $\beta_T$ relative to the default settings, for our annealing procedures on $100$ randomly generated RAN1 problems at each of 3 sizes: C2 (32 variables), C4 (127 variables), and C12 (1100 variables). Results for AC3 are presented in supplementary materials.

\begin{figure}[htb]
  \centering
    \includegraphics[width=0.4\textwidth]{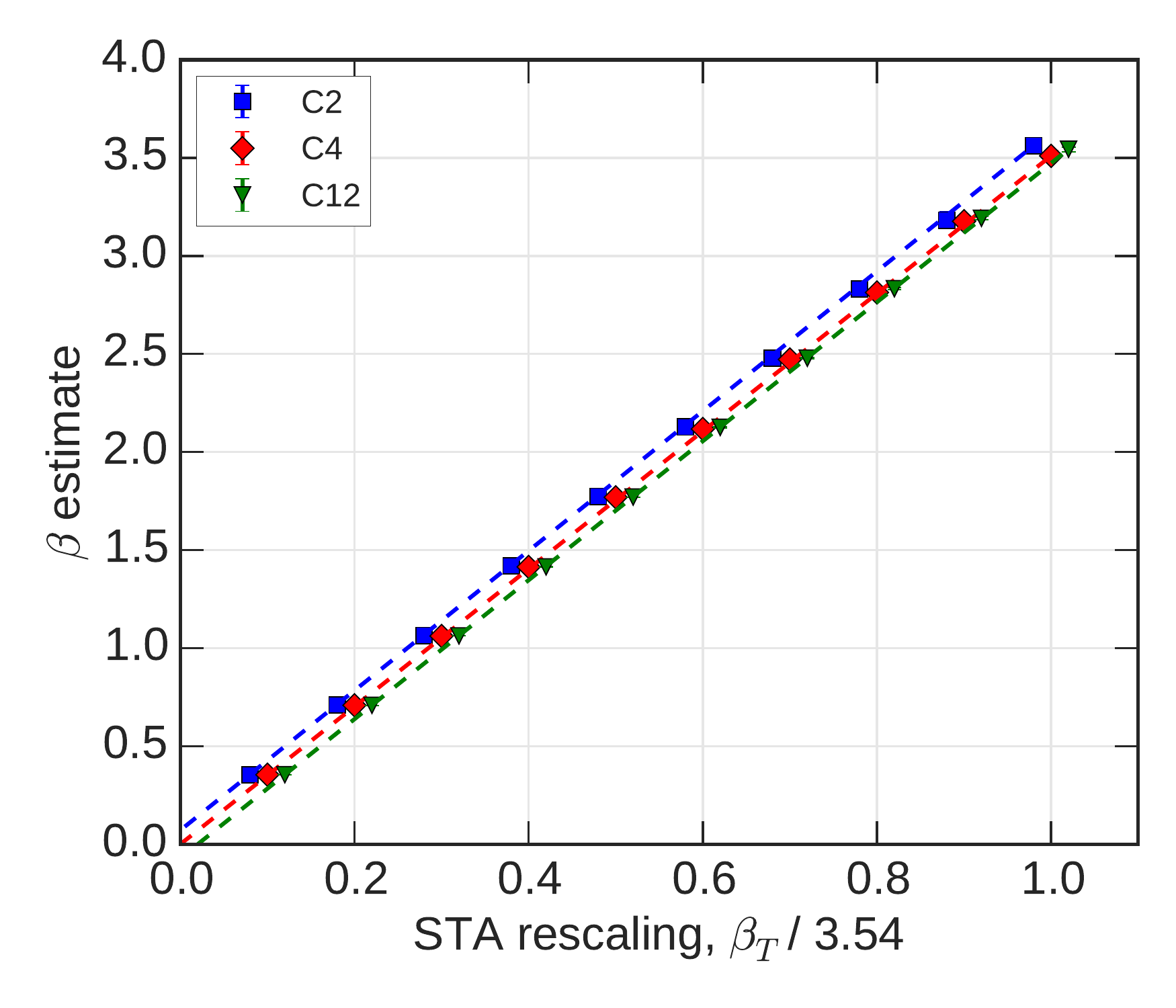}
    \includegraphics[width=0.4\textwidth]{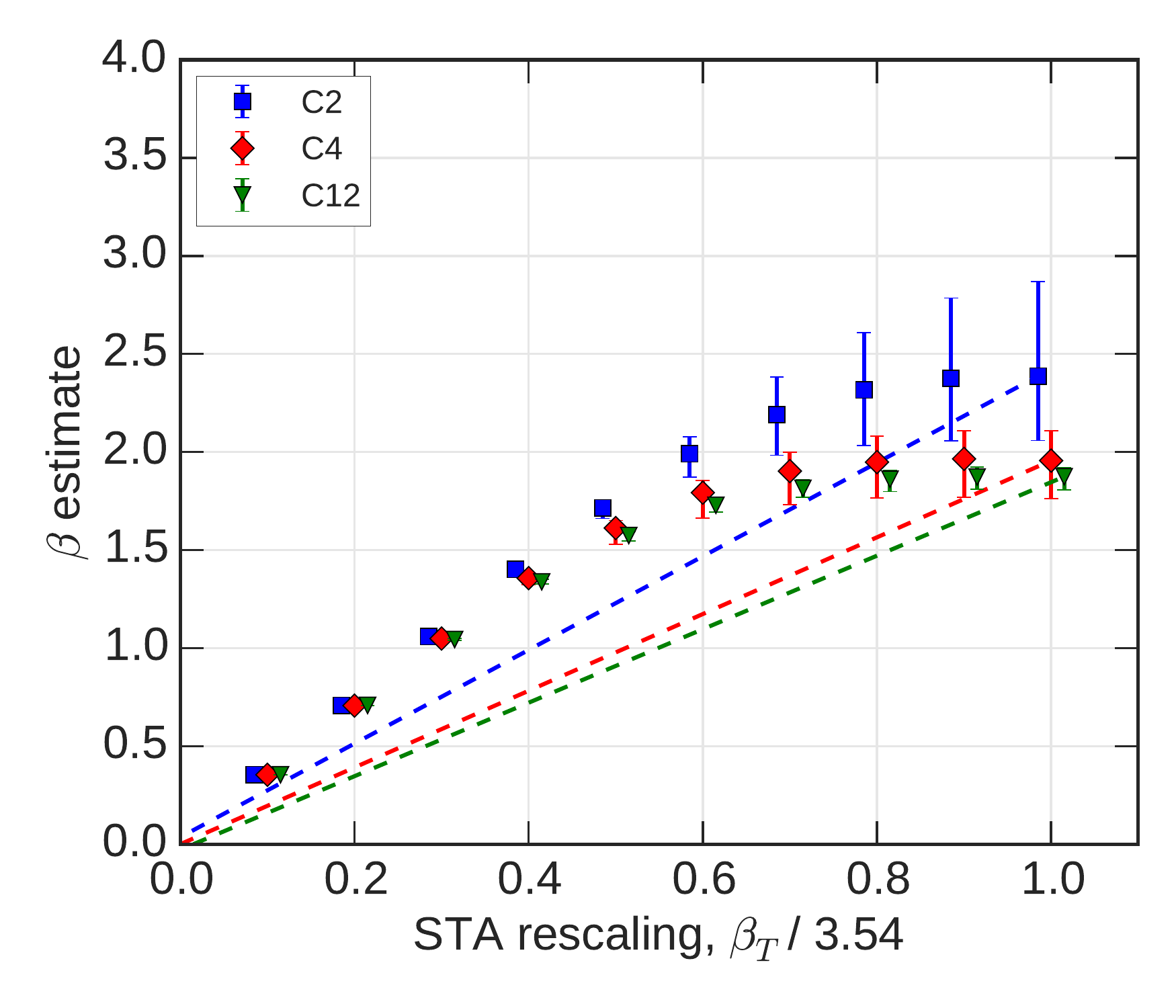}
    \caption{\label{fig:rescalingSA} Bars represent quartiles over $100$ instances of RAN1 at each scale, estimates for $\beta$ for in each case $10^4$ samples generated by the STA. C2 and C12 data points are offset for visibility. (left) Temperature estimates by the MLPL method. The MLPL estimate for thermal annealers matches the terminal model at all scales. (right) Temperature estimates by ML. A non-linear trend is apparent due to ergodicity breaking, which is not captured by MLPL estimates. }
\end{figure}

We first consider the behavior of the STA with a range of terminal inverse temperatures between 0 and the default value for the problem class. Due to ergodicity breaking, we expect samples to fall out of equilibrium before the terminal temperature is reached, and so the distribution may be characterized by a value of $\beta$ smaller than $\beta_T$, reflecting the range of inverse temperature for which dynamics slowed down. Figure \ref{fig:rescalingSA} shows the maximum likelihood and MLPL estimates based upon the same sample sets. We see that the MLPL estimates follow a linear curve, which would indicate that the terminal model is indeed the best fit to the samples, with no evidence of the ergodicity breaking we describe. By contrast, the maximum likelihood estimator is concave, with $\beta$ significantly smaller than $\beta_T$. 

\begin{figure}[htb]
  \centering
    \includegraphics[width=0.4\textwidth]{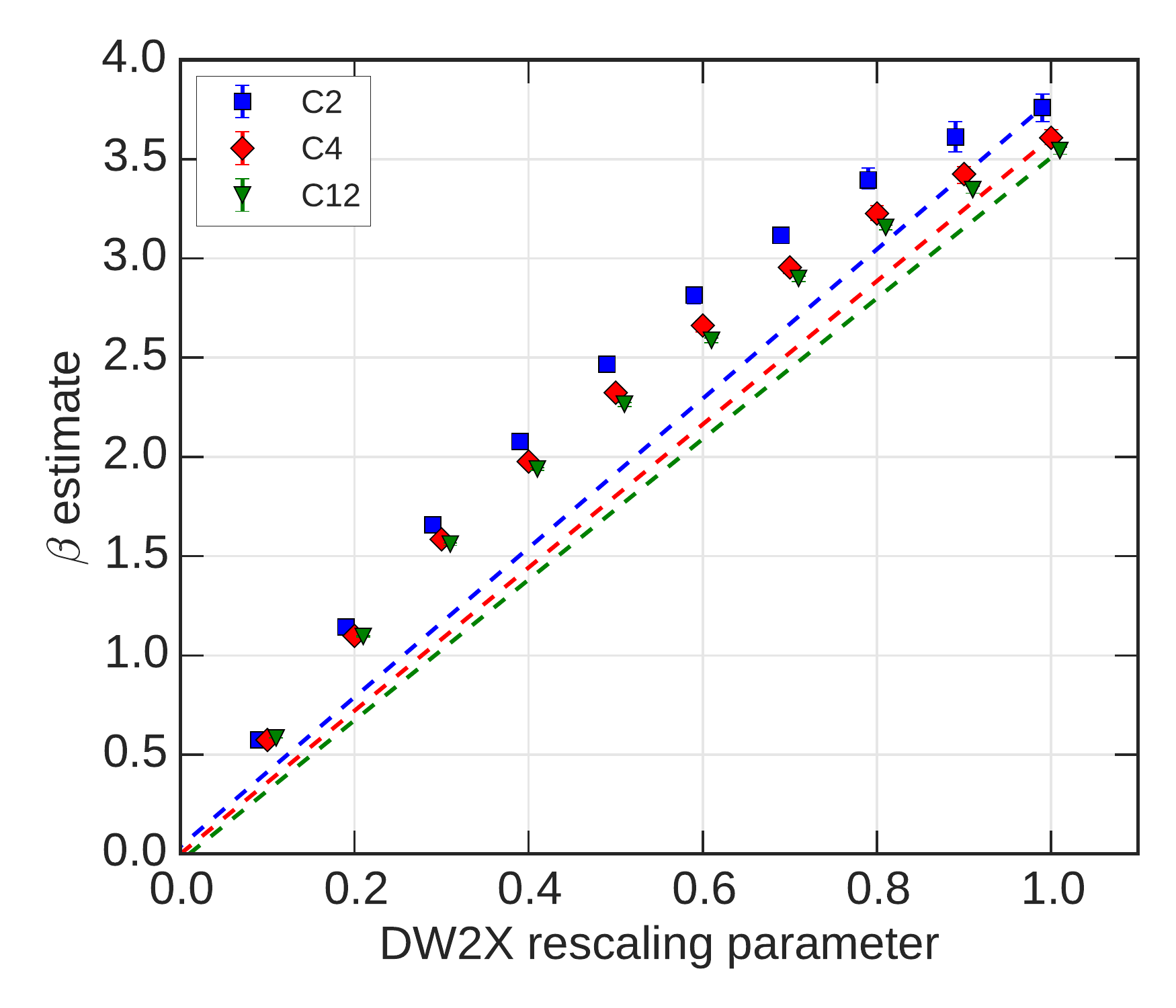}
    \includegraphics[width=0.4\textwidth]{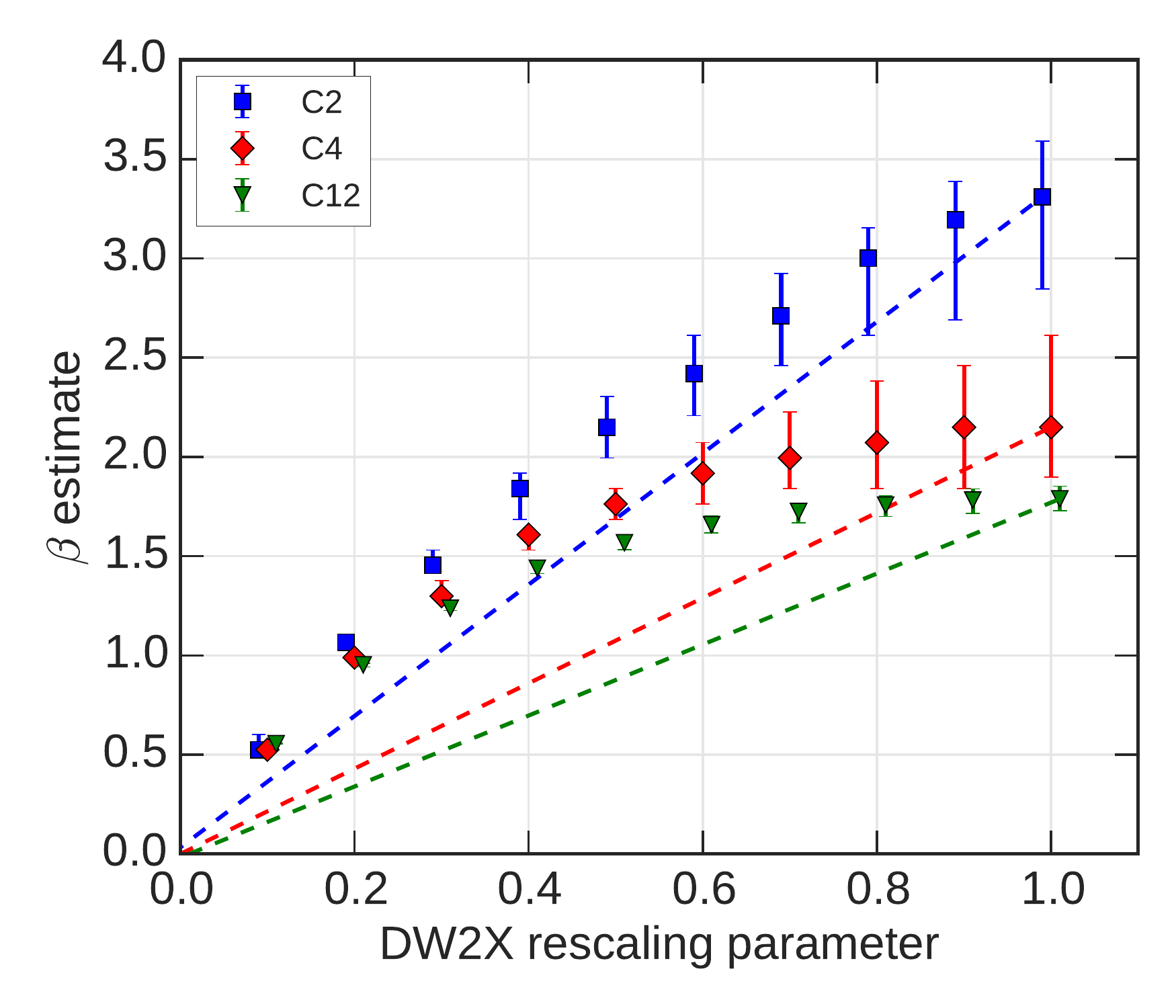}
    \caption{\label{fig:rescalingDW2X} As per Figure \ref{fig:rescalingSA}, but using samples from the DW2X. (left) The MLPL method is non-linear, unlike the STA results. (right) The maximum likelihood method shows qualitatively similar features to the STA. Note that the C12 result for the MLPL estimate at $r=1$ is identical to that in Figure \ref{fig:rescalingSA}, the STA parameter $\beta_T$ was chosen to meet this criteria as discussed in Section \ref{sec:BlockedGibbs}. }
\end{figure}

The DW2X can also be manipulated by changing the rescaling parameter $r$ on the interval $[0,1]$. We thus repeat this experiment using sample sets from the DW2X. In Figure \ref{fig:rescalingDW2X} we see that both locally (MLPL) and globally (ML) the estimators are concave as a function of this rescaling. As with the STA, estimates are consistently larger for MLPL than maximum likelihood, and maximum likelihood estimates decrease for larger, more complicated, problems. Maximum likelihood exhibits some small decrease with system size. A naive interpretation of the rescaling parameter might lead to a general hypothesis that $\beta \propto r$, where the constant of proportionality can be determined by the terminal energy scale in annealing (see Figure \ref{fig:schedule}). However, the physical dynamics of qubits are controlled by a {\em single qubit freeze-out} phenomenon that is discussed in supplementary material, with data summarized in Figure \ref{fig:rescalingDW2X1bit}. The single qubit freeze-out figure implies that the non-linearity of the MLPL curve in spin-glass problems is a function of the problem precision (granularity of the settings of $J_{ij}$ and $h_i$), or more specifically, the pattern of energy gaps. Problems of higher precision, such as the AC3 problem class, have less pronounced non-linearity. The single qubit freeze-out phenomenon also anticipates the system size dependence seen in Figure \ref{fig:rescalingDW2X}, and in some other models not presented. AC3 results, and further discussion of this point, are in supplementary materials.

We believe the phenomenon underlying both MLPL and maximum likelihood to be easily understood, and similar in both the DW2X and STA, although the interaction of the dynamics with the energy landscape may be quite different. Figure \ref{fig:schematic} provides a useful example to explain this. When ergodicity is broken, and the sample set is divided over subspaces, each subset relaxes towards the terminal model restricted to the subspace; the MLPL method effectively averages an estimate over these subspaces. By contrast the maximum likelihood estimate accounts in part for the distribution between modes, characterized at an early (higher energy) stage of the anneal that is better described by smaller inverse temperature. Note that the equilibrated distribution at the ergodicity breaking point is a quantum one for the DW2X, unlike thermal annealers, so we ought to understand the deviation in the local and global temperatures in terms of the quantum parameterization ($\Delta(t),E(t)$)~\cite{Amin:SQS}. Still, the classical description in terms of $\beta$ (\ref{eq:PBoltzmann}) provides the correct intuition. 

Response curves such as these can be used to choose a suitable parameterization of the terminal model -- we can choose $\beta_T$ so as to minimize KL-divergence. Similar curves could be constructed for any parameter, and with a distribution that is subject to post-processing. In the absence of the maximum likelihood curve (due to absense of approximates to the energy), a local information curve such as MLPL curve can be used as a compromize. The maximum likelihood curves, and MLPL curve for the DW2X, are problem dependent. Some temperature estimators require knowledge of, or assumptions about, the form of these curves -- notably the method of Benedetti et al.~\cite{PerdomoOrtiz:EET} and our multi-canonical method, both described in supplementary materials.

To judge quality of the approximation at the ``best'' $\beta$, it is appropriate to consider the quantity being minimized, KL-divergence. However, this is a difficult quantity to estimate from samples in the absence of parametric assumptions. We present a method for estimation of KL-divergence in supplementary materials, but find that for RAN1 and AC3 problems it is ineffective at C12 scales. This is an important reason to consider an alternative such as the MSE estimator.

\subsection{The mean square error estimator}
\label{sec:MSEexperiments}

\begin{figure}[htb]
  \centering
    \includegraphics[width=0.45\textwidth]{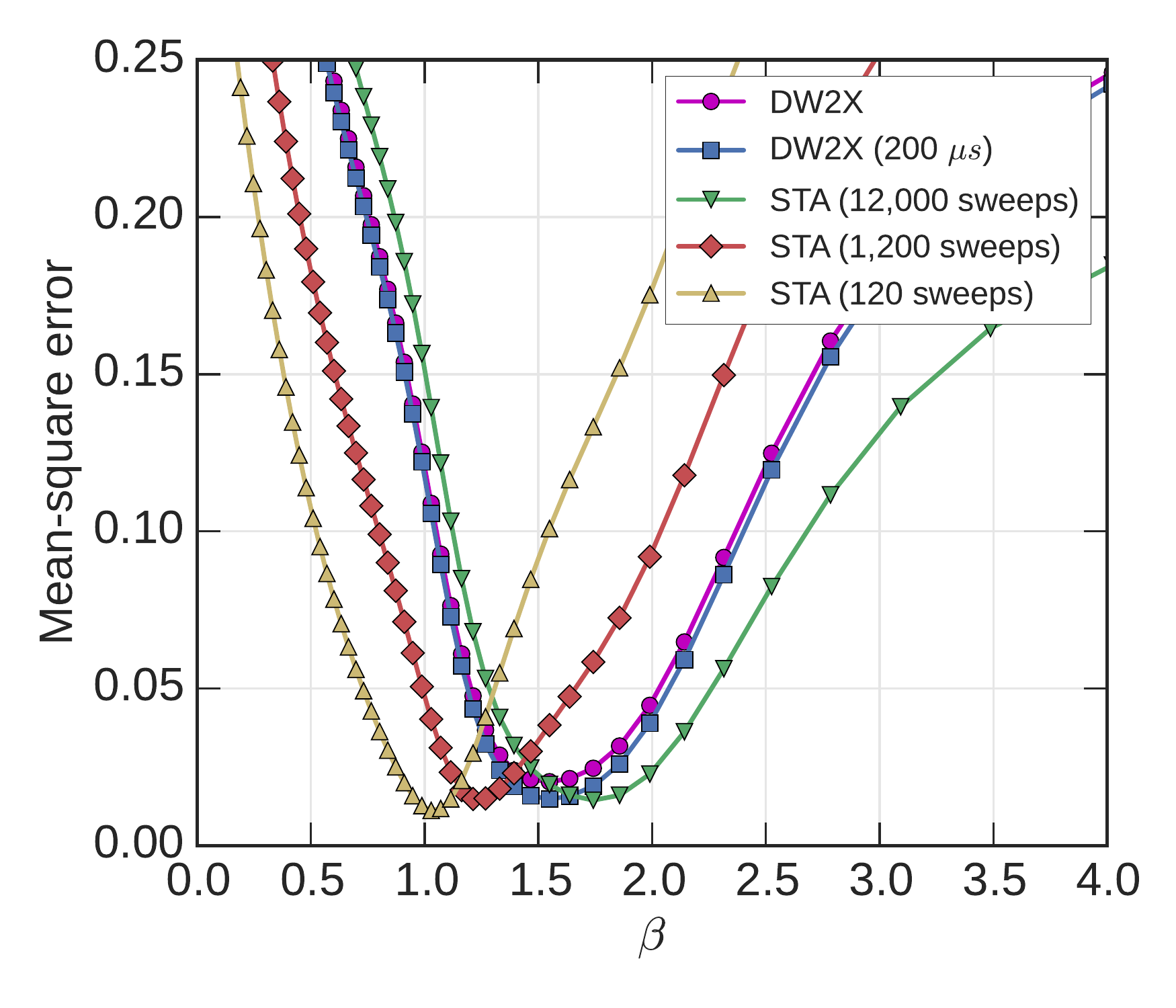}
    \includegraphics[width=0.45\textwidth]{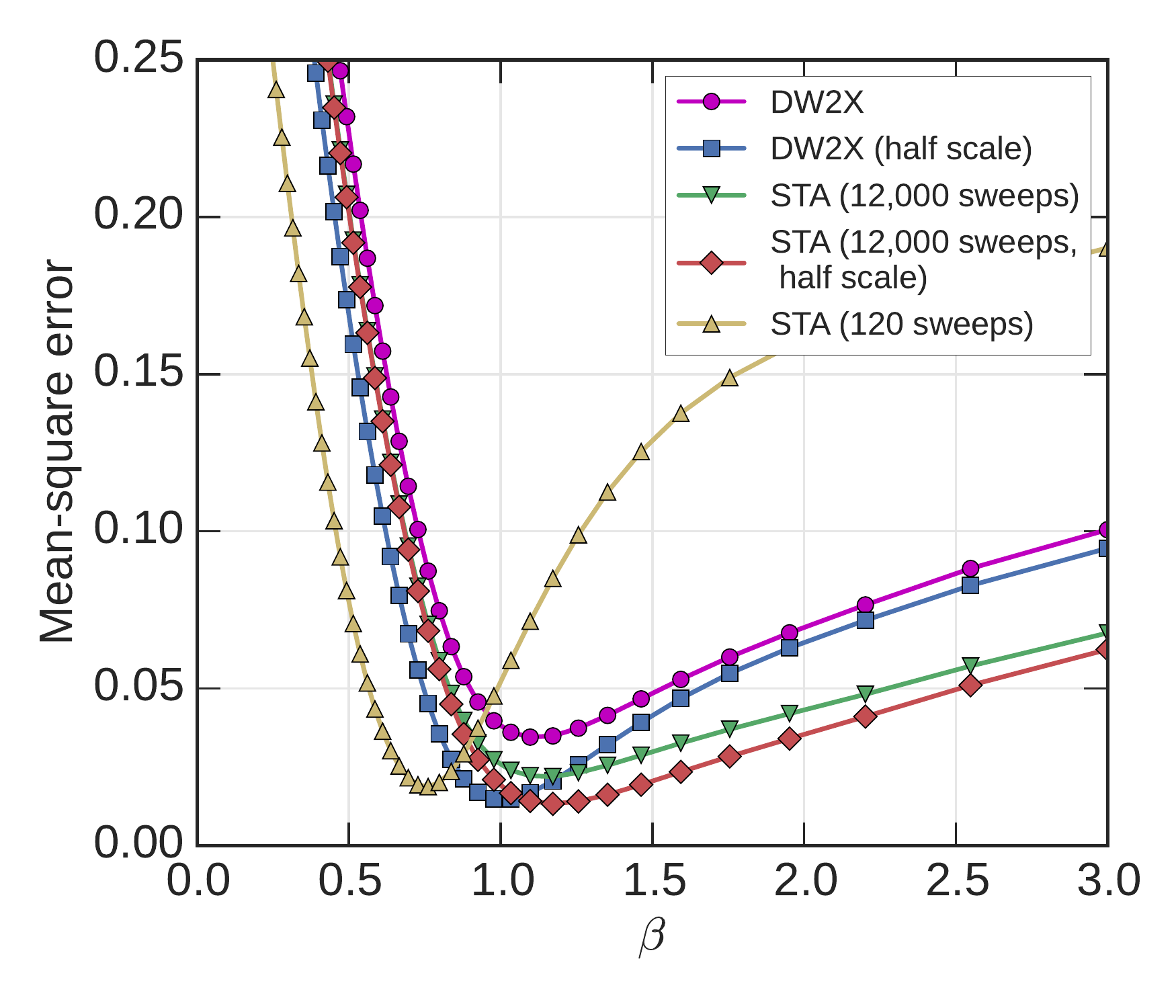}
    \caption{\label{fig:MSE} Both the DW2X and STA can be used to sample from a RAN1 problem with small errors over an intermediate range of $\beta$. Objective performance is shown for two typical instances under several annealer operating conditions. (left) Results for the AC3 exemplar. (right) Results for the RAN1 exemplar. Standard errors determined by jack-knife methods are negligible compared to the marker size. To avoid clutter we show only variation of the anneal time in the left figure, and only variation of the rescaling parameter ($r=0.5$ in the DW2X, $\beta_T/2$ in the STA) in the right figure. }
\end{figure}

The mean square errors on correlations associated to the DW2X and STA for a typical instance of RAN1, and a typical instance of AC3, are shown for the C12 (1100 variables) problem size in Figure \ref{fig:MSE}. Though there is significant variation between the curves associated to different instances of these models, we chose amongst 100 random instances exemplars that are typical in the minimizing temperature and the mean square error $(\beta, \mathrm{MSE})$ for DW2X. 

Both the DW2X and STA performance is best characterized at intermediate $\beta$ values, and variation from the default annealer settings is also shown to modify MSE. In the RAN1 exemplar we demonstrate the effect of halving the terminal temperature or rescaling parameter, which improves MSE (except perhaps at very large values for $\beta$). In the case of the AC3 exemplar we demonstrate the effect of varying the anneal time. Results indicate that longer anneals tend to improve MSE at lower temperature. To prevent clutter we have shown only one variation of a default parameter per model. Switching the parameter being varied results are qualitatively similar. Annealing for longer is expected to allow equilibration to lower temperatures, and so a better match is to be expected. Annealing with smaller $r$ or $\beta_T$, concentrates annealing resources towards the initial part of the anneal where dynamics are effective (rather than at the end of the anneal where ergodicity breaking has already occurred and cannot be mitigated). It also reduces the tendency of samples to settle towards local minima that might distort the approximation for intermediate $\beta$. In the case of the DW2X, some noise sources and quantum mechanical phenomena can complicate this simple picture, but this is not obviously at play.

\begin{figure}[htb]
  \centering
    \includegraphics[width=0.45\textwidth]{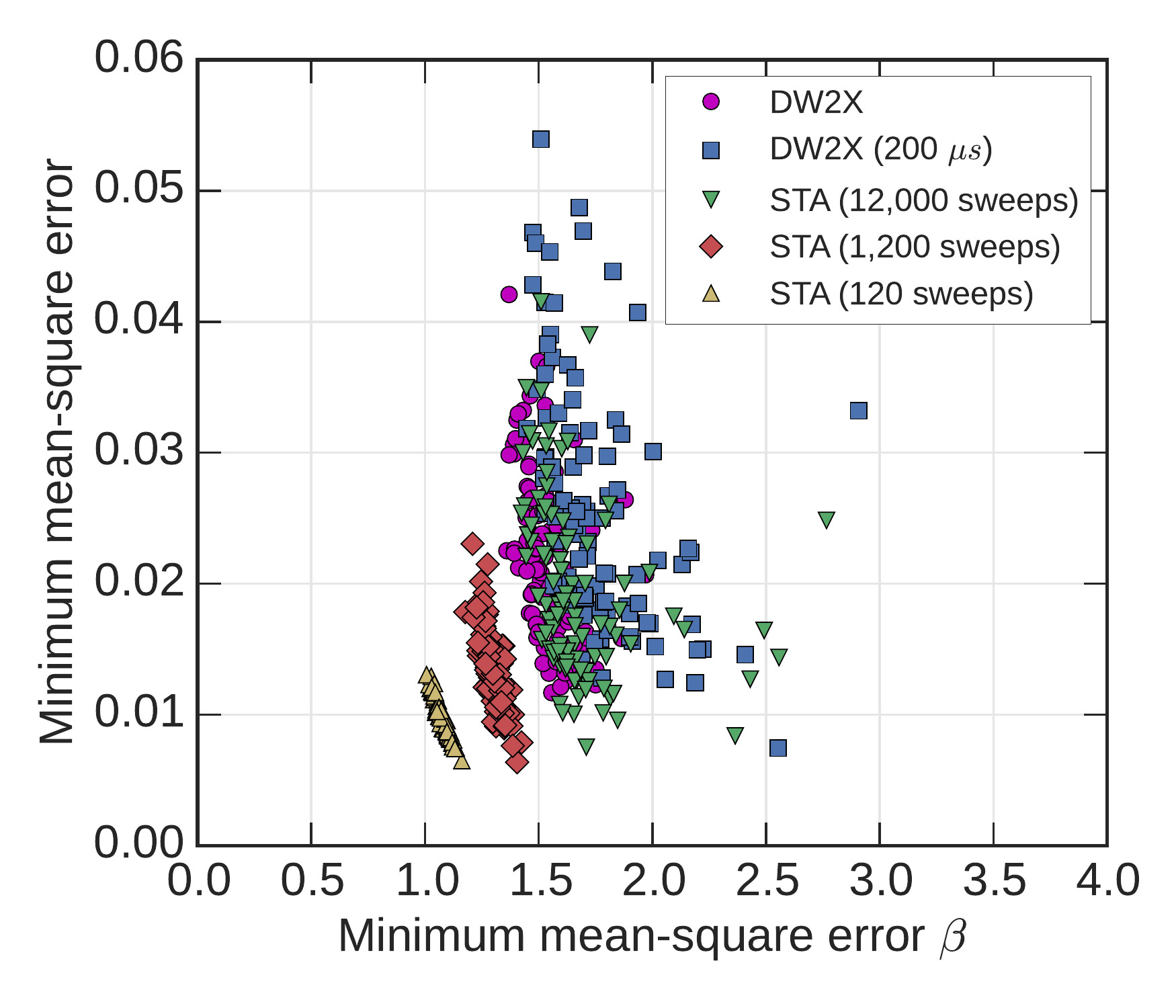}
    \includegraphics[width=0.45\textwidth]{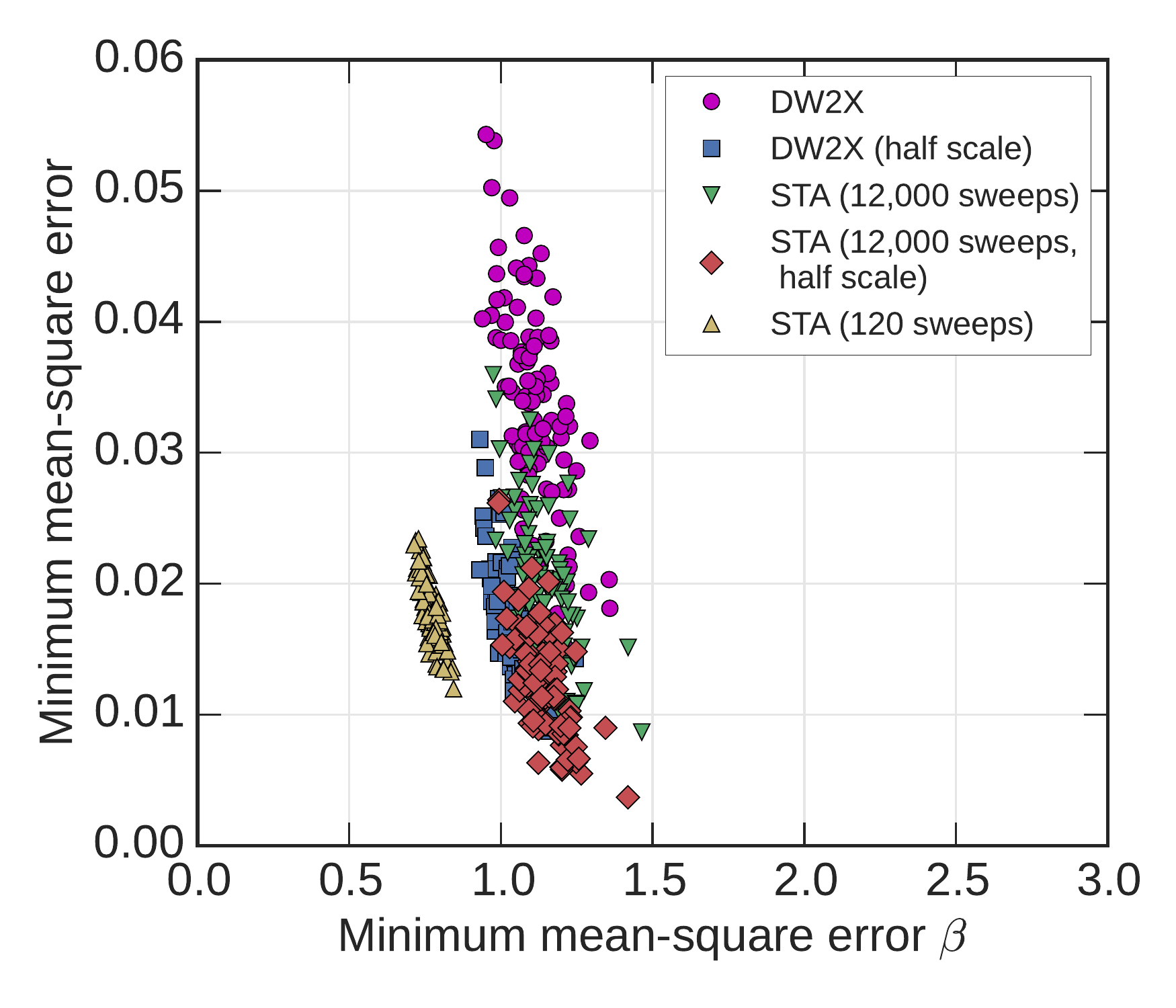}
    \caption{\label{fig:EstPerEst2} Statistics over the set $100$ RAN1 (left) and AC3 (right) problems at C12 scales. Small objective values at large inverse temperatures are difficult to obtain, and so desirable in a heuristic sampler. Sampling effectively at small inverse temperature is less valuable (e.g., 120 sweep STA). Modification of the annealing parameters significantly changes the distribution, allowing more effective emulation at some inverse temperatures. To avoid clutter we show only variation of the anneal time in the left figure, and only variation of the terminal model rescaling ($r$, $\beta_T$) in the right figure; the effects are qualitatively similar in each of these models at C12 scale.}
\end{figure}

\begin{figure}[htb]
  \centering
    \includegraphics[width=0.45\textwidth]{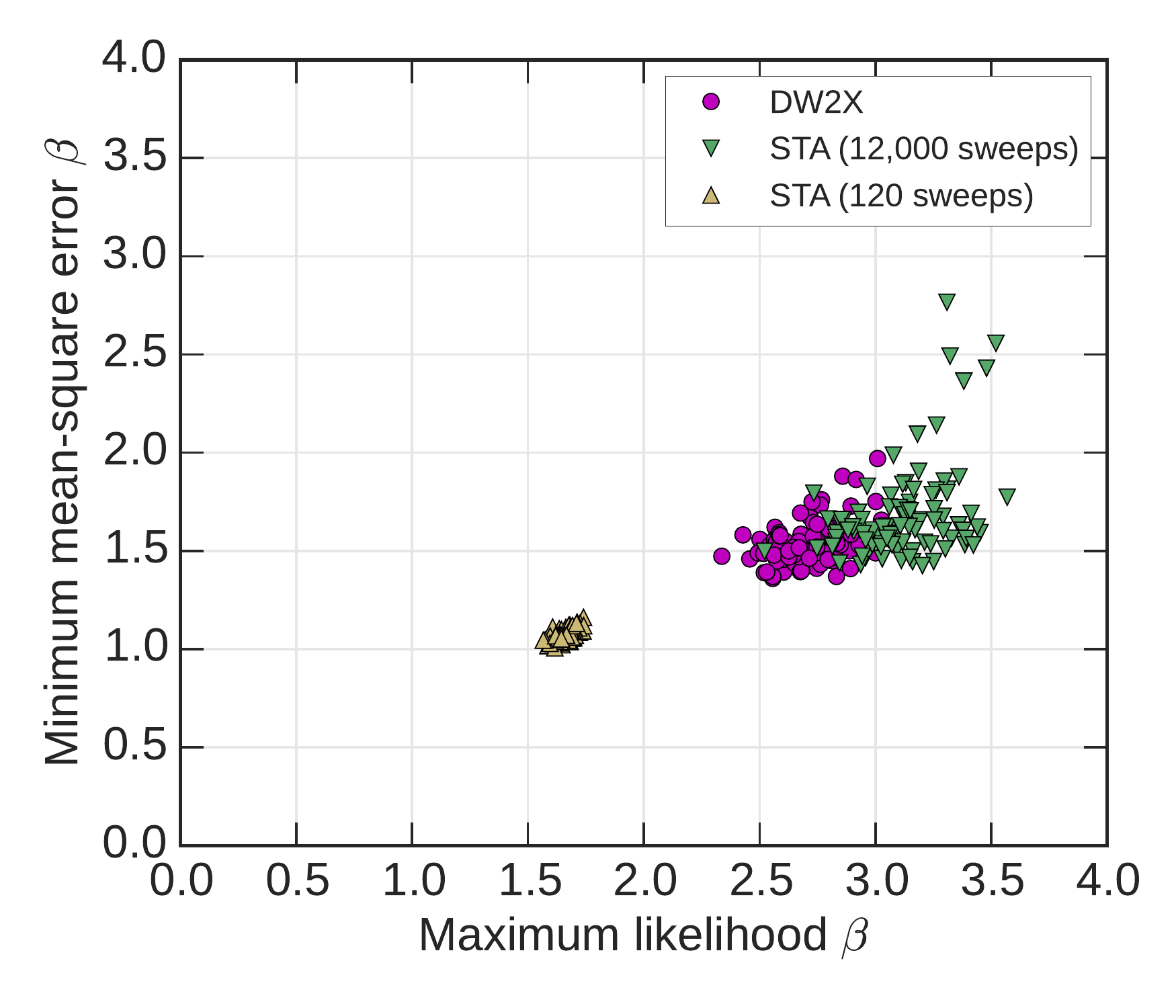}
    \includegraphics[width=0.45\textwidth]{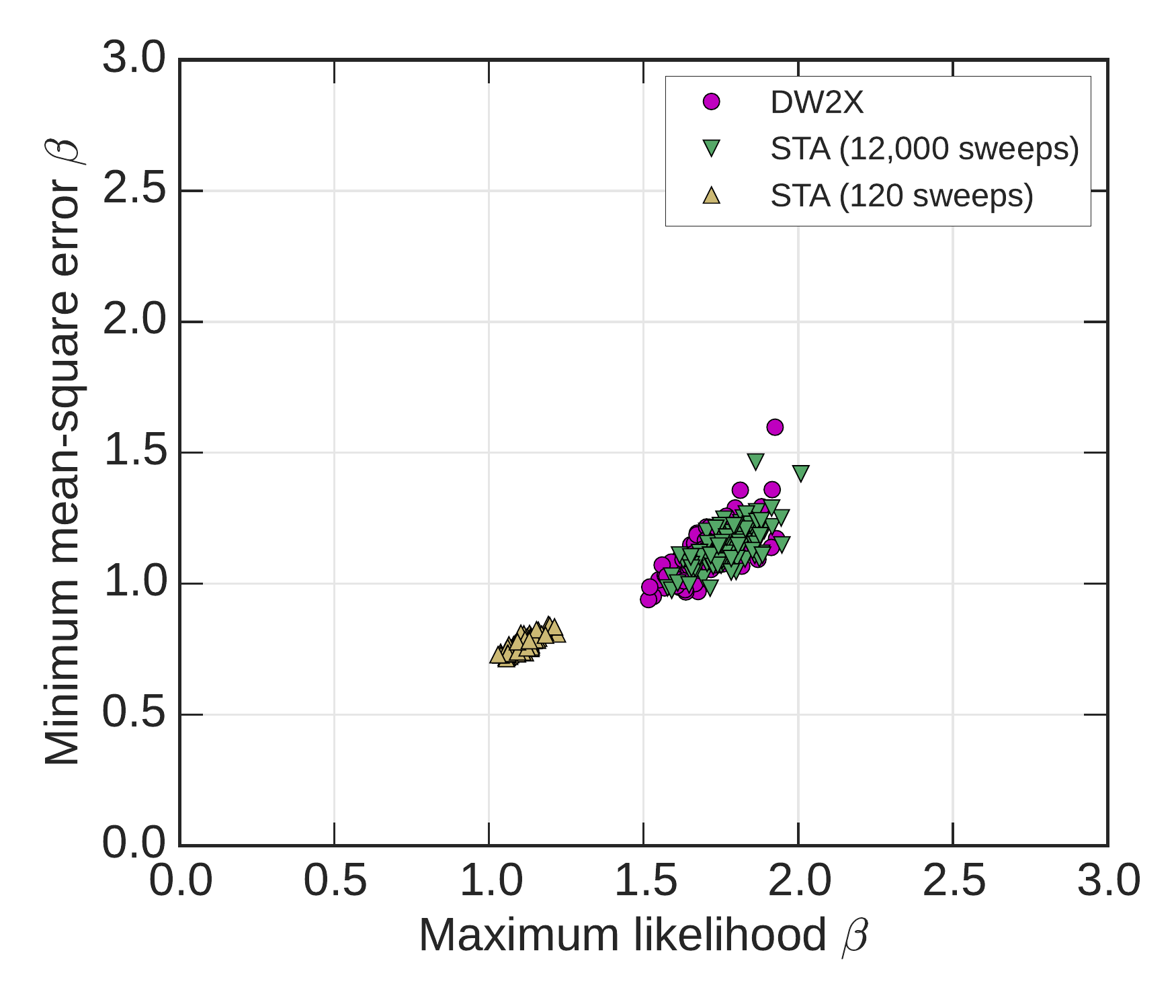}
    \caption{\label{fig:EstPerEst} Statistics over the set $100$ AC3 (left) and RAN1 (right) problems at C12 scales, as per Figure \ref{fig:EstPerEst2}. Minimum MSE and maximum likelihood estimators of temperature give different, but strongly correlated, results. The maximum likelihood estimate is typically larger, a partial explanation is the sinking of samples towards local minima late in the anneal, which through its impacts on mean energy has consequences for the maximum likelihood estimate. }
\end{figure}

Figure \ref{fig:EstPerEst2} shows the mean square error achieved against the point it is minimized, for a large set of C12 (1100 variable) problem instances. Ideally we wish for heuristics that are both sampling at large $\beta$ and with small errors. Figure \ref{fig:EstPerEst} shows the minimum MSE estimator in comparison to the maximum likelihood estimator. These two estimates indicate different operational temperature ranges. The fact that the maximum likelihood estimator is significantly larger is to be expected in annealers. Late in the anneal samples sink towards their respective local minima, decreasing the mean energy significantly (see Figure \ref{fig:schematic}). The mean energy is strongly dependent on the local relaxation, and hence the local inverse temperature, which as we saw in the previous section is large. In the case of MSE, the correlation error is for most models not strongly affected by this local decrease in energy. It is more sensitive to the distribution between modes, which is set only by the temperatures characterizing ergodicity breaking. 

By annealing with modified $\beta_T$ or $r$, improvements are made for sampling intermediate or small $\beta$. By contrast, it is relatively hard to sample effectively from large values of $\beta$ by modifying these parameters; additional time resources are required to make an impact. For this reason, we may argue that, generally, the larger the inverse temperature estimate, the more useful the annealer will be for hard sampling applications. However, it is important to note that an estimate for $\beta$, independent of the objective measure, may be risky or misleading. In Figure \ref{fig:EstPerEst}(right), the DW2X system at full scale indicates a lower minimum MSE $\beta$ than the DW2X system operating at half scale. However, we can see that at this larger $\beta$ value the half scale system is still more effective as a heuristic. 

\subsection{Effectiveness of post-processing}
\label{sec:postProcessing}

In Sections \ref{sec:MLandMLPL} and \ref{sec:MSEexperiments} we demonstrated how adjusting anneal duration, or the terminal temperature, can allow better objective outcomes. In this section we consider briefly the effect of post-processing by one sweep of blocked Gibbs as discussed in Section \ref{sec:PP}. This post-processing changes dramatically the local distribution of samples, hence the MLPL estimate. However, the KL-divergence and mean square error, and the temperatures minimizing these objectives, are affected by a combination of the local and global distribution and so are modified in a non-trivial way by post-processing. Post-processing always strongly modifies local distribution properties, but only in the easy to sample regime (at small $\beta$) does it significantly impacts global distribution problems.

\begin{figure}[htb]
  \centering
    \includegraphics[width=0.45\textwidth]{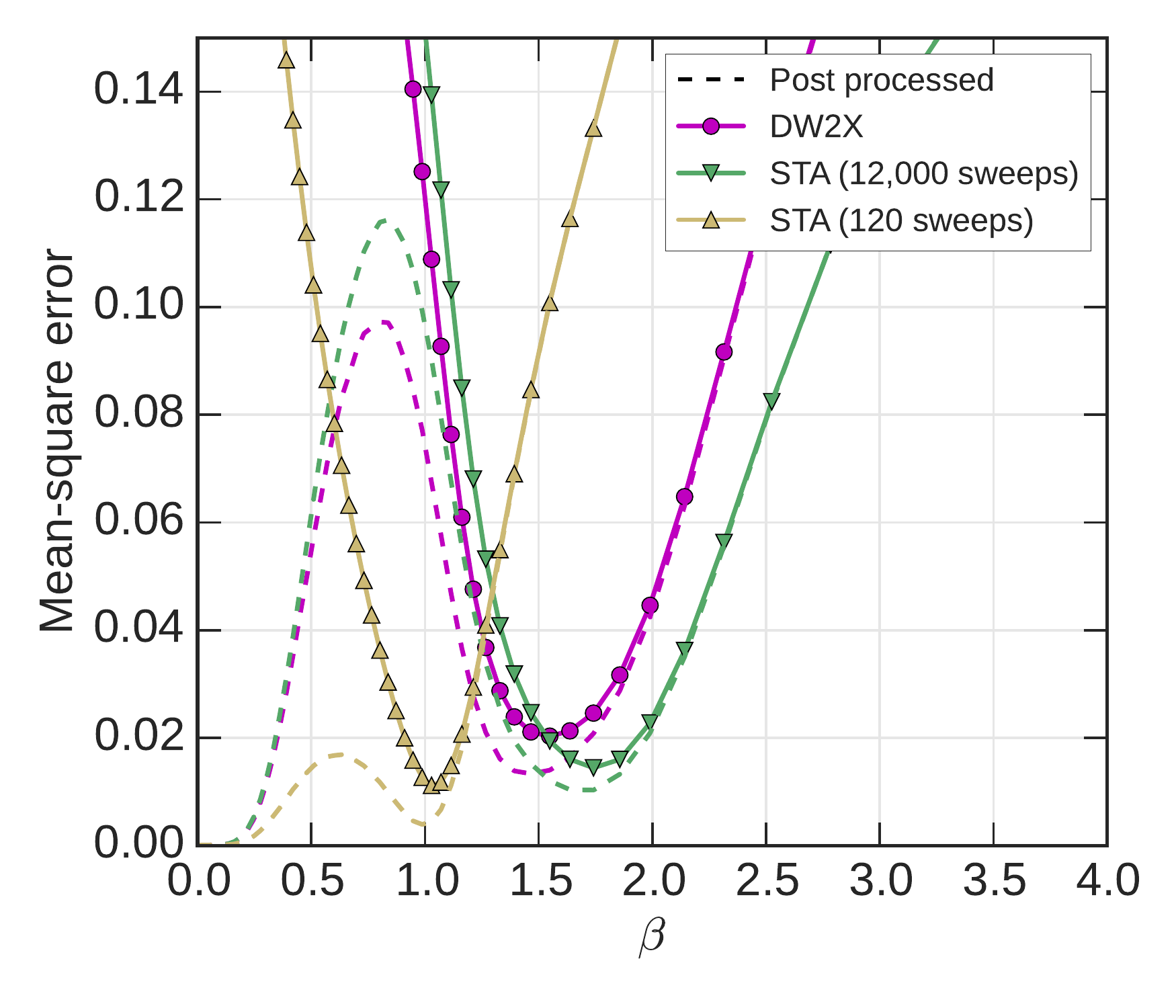}
    \includegraphics[width=0.45\textwidth]{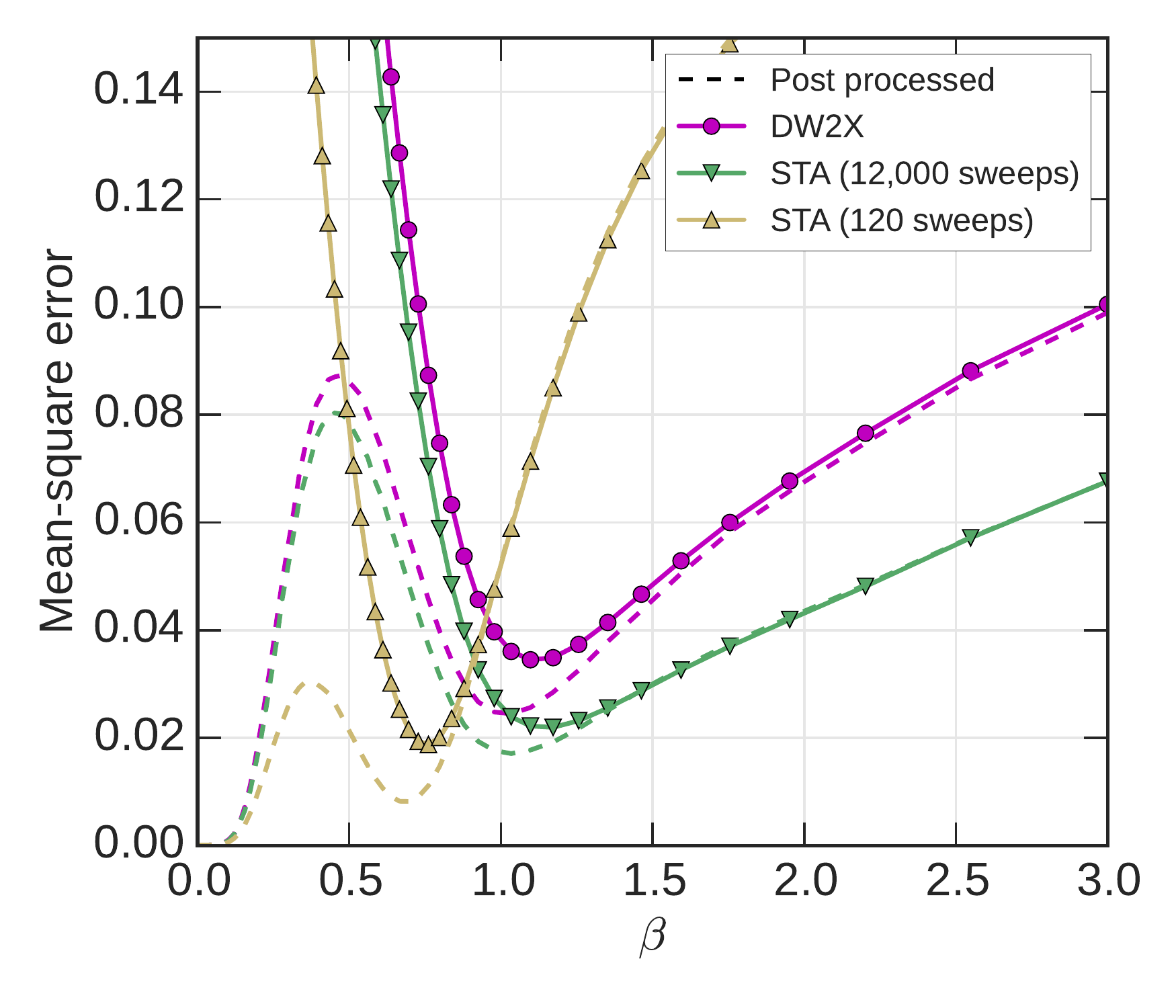}
    \caption{\label{fig:MSE2} Simple forms of post-processing can have a quantitatively large effect on objectives at small and intermediate temperatures. The C12 problem exemplars consistent with Figure \ref{fig:MSE} are shown. At low $\beta$ a single sweep of blocked Gibbs can completely correct all errors. At high $\beta$ there is relatively little effect, however the effect is significant in the intermediate range of inverse temperature where the annealers can be considered most effective.}
\end{figure}

Figure \ref{fig:MSE2} shows that MSE on correlations are, as expected, improved by post-processing. The improvements are dramatic in the regime $\beta\approx 0$, impressive over intermediate values of $\beta$, but almost negligible for larger $\beta$. 
After post-processing, the MSE curve has two minima. There is no guarantee there should be two local minima working with arbitrary distributions. The first local minimum ($\beta=0$) is evidence for the power of post-processing. Since one sweep of blocked Gibbs samples effectively at $\beta \approx 0$, independent of the initial condition $P_A$, $\beta=0$ will be a global minimum for any heuristic distribution. The second local minimum appears due to the closeness (at the macroscopic level) of the annealing distribution to some particular low-temperature Boltzmann distribution, a sweet spot of operation that may be of practical interest. 

Post-processing reduces the error everywhere, but more so at smaller $\beta$ where the time-scales for macroscopic redistribution are shorter. For this reason we expect the minimizing value of $\beta$ to move leftward with post-processing. If the post-processing allows global redistribution of samples we may anticipate the disappearance of the local maximum separating the ``easy for post-processing regime'' from the ``good for this annealer'' regime; at which point a best operational regime for the annealer is less clear. However, we can assume that powerful post-processing of this kind is too expensive in the types of multi-modal problems where annealers are useful.

\begin{figure}[htb]
  \centering
    \includegraphics[width=0.45\textwidth]{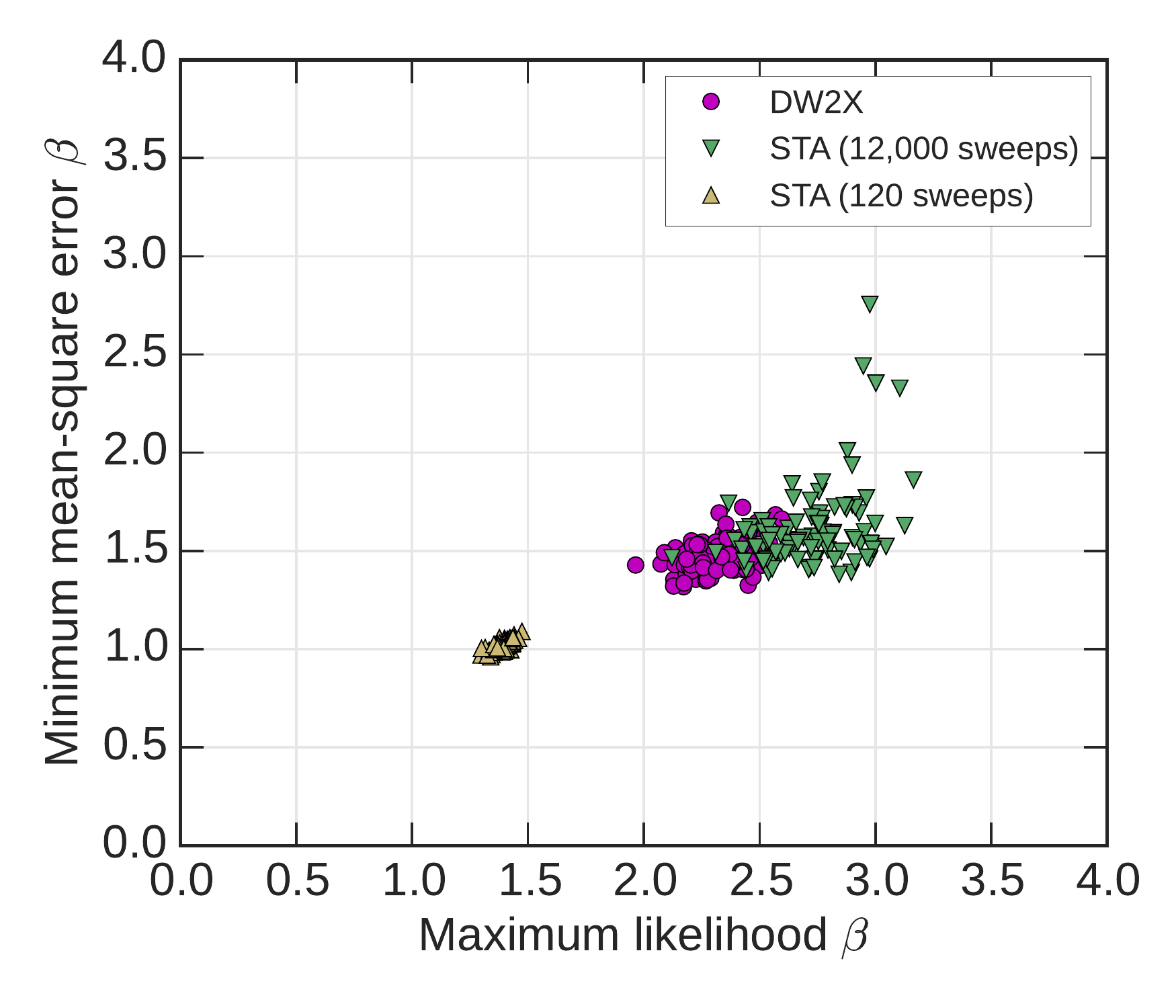}
    \includegraphics[width=0.45\textwidth]{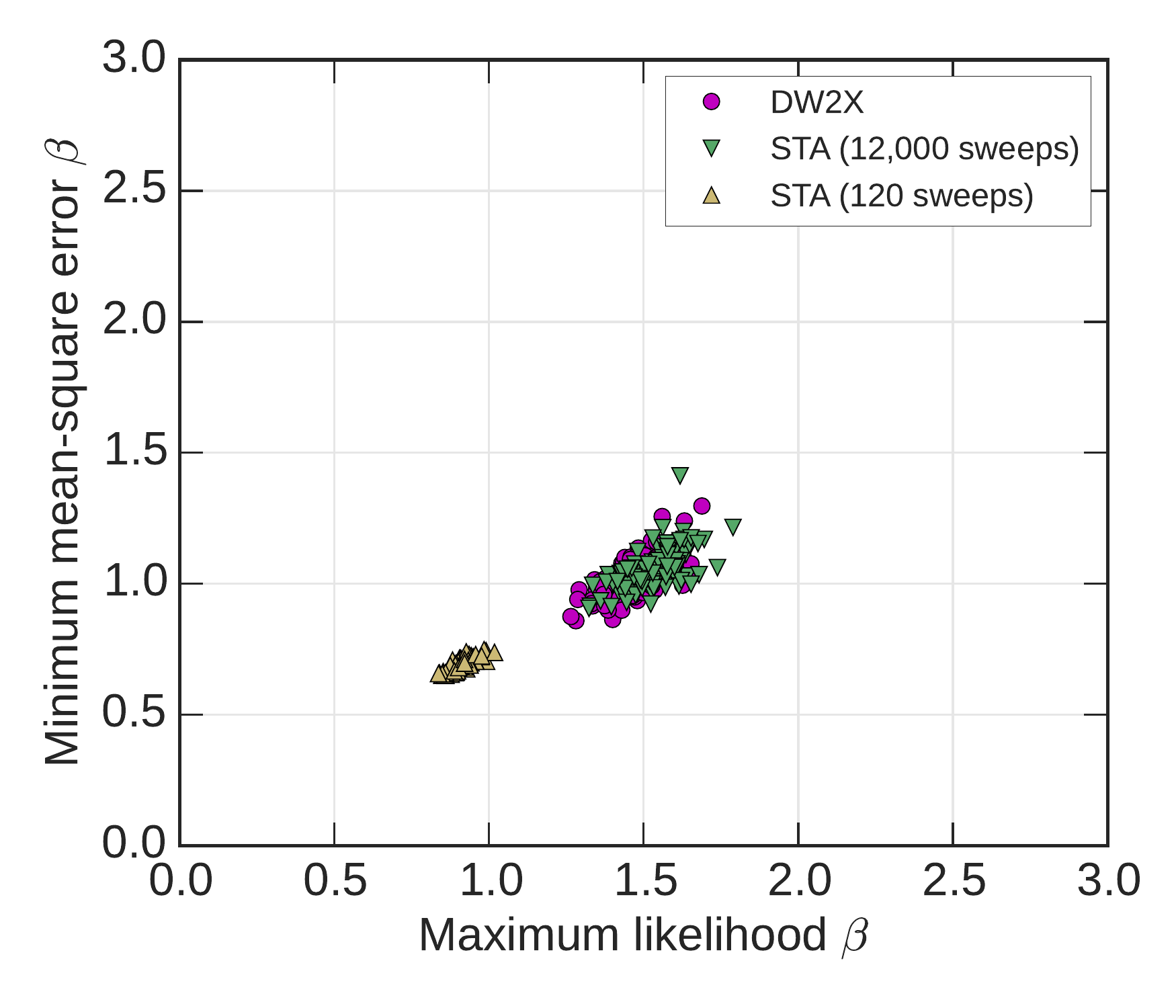}
    \caption{\label{fig:EstPerEst_PP} As per Figure \ref{fig:EstPerEst}, but now all distributions are modified by one sweep of blocked Gibbs sampling. The estimators for inverse temperature are reduced, as the effect of the local distribution (characterized by larger inverse temperature) is partially removed.}
\end{figure}

Figure \ref{fig:EstPerEst_PP} shows the statistics for the local minimum mean square error estimator, and its relation to the local maximum likelihood estimator, to be compared against Figure \ref{fig:EstPerEst} that has no post-processing. The local minimizer is the right most local minimum of the post-processed curve (see Figure \ref{fig:MSE2}), indicating the good operating regime. A global minimum is always at $\beta=0$, but this is not of interest as it reflect the post-processor and not the heuristic.

In Figure \ref{fig:EstPerEst_PP}(left) we see that, relative to the distribution without post-processing, there is a slight shift leftwards in all distributions, and significant shift downwards that appears approximately proportional to the MSE without post-processing. The effect of local relaxation is to give the impression of better estimation at low temperature. Here the effect is partially lifted to reveal that the macroscopic distribution may be characterized by slightly smaller $\beta$.

In Figure \ref{fig:EstPerEst_PP}(right) we see how the minimum MSE estimate decreases in all annealers, but by a less significant amount than the downward trend in the maximum likelihood estimator. The two remain strongly correlated under post-processing. It is natural to expect that the local relaxation, which shifts samples towards their local minimum, may have a bigger impact on KL-divergence than MSE, because it is easy to raise the energy by post-processing which impacts maximum likelihood in a systematic manner, but it is difficult to redistribute samples macroscopically, which may be required to alter minimum MSE.

The particular effects demonstrated on the exemplar instances of Figure \ref{fig:EstPerEst_PP} are reflected at the distribution level in Figure \ref{fig:EstPerEst2_PP}.

\begin{figure}[htb]
 \centering
  \includegraphics[width=0.45\textwidth]{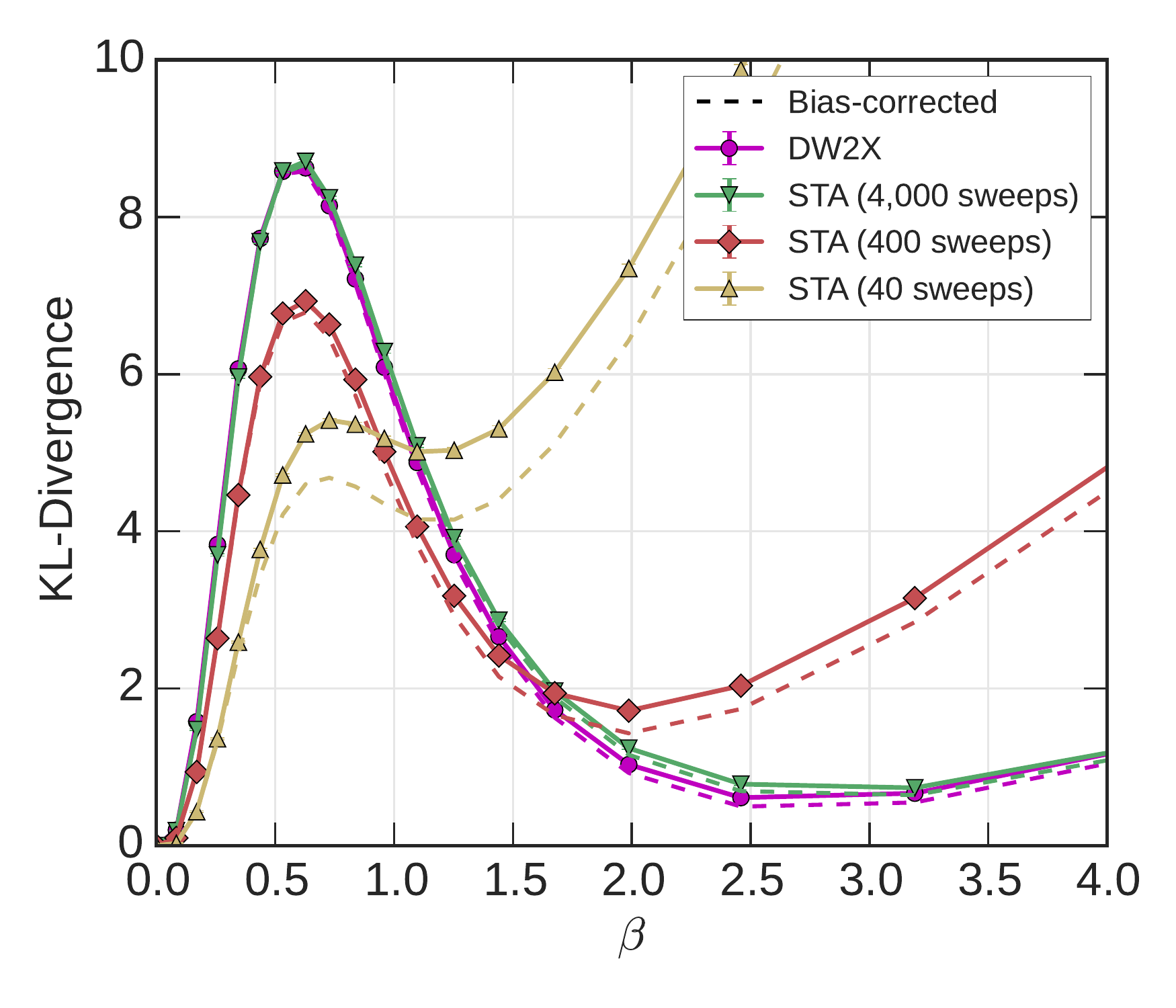}
  \includegraphics[width=0.45\textwidth]{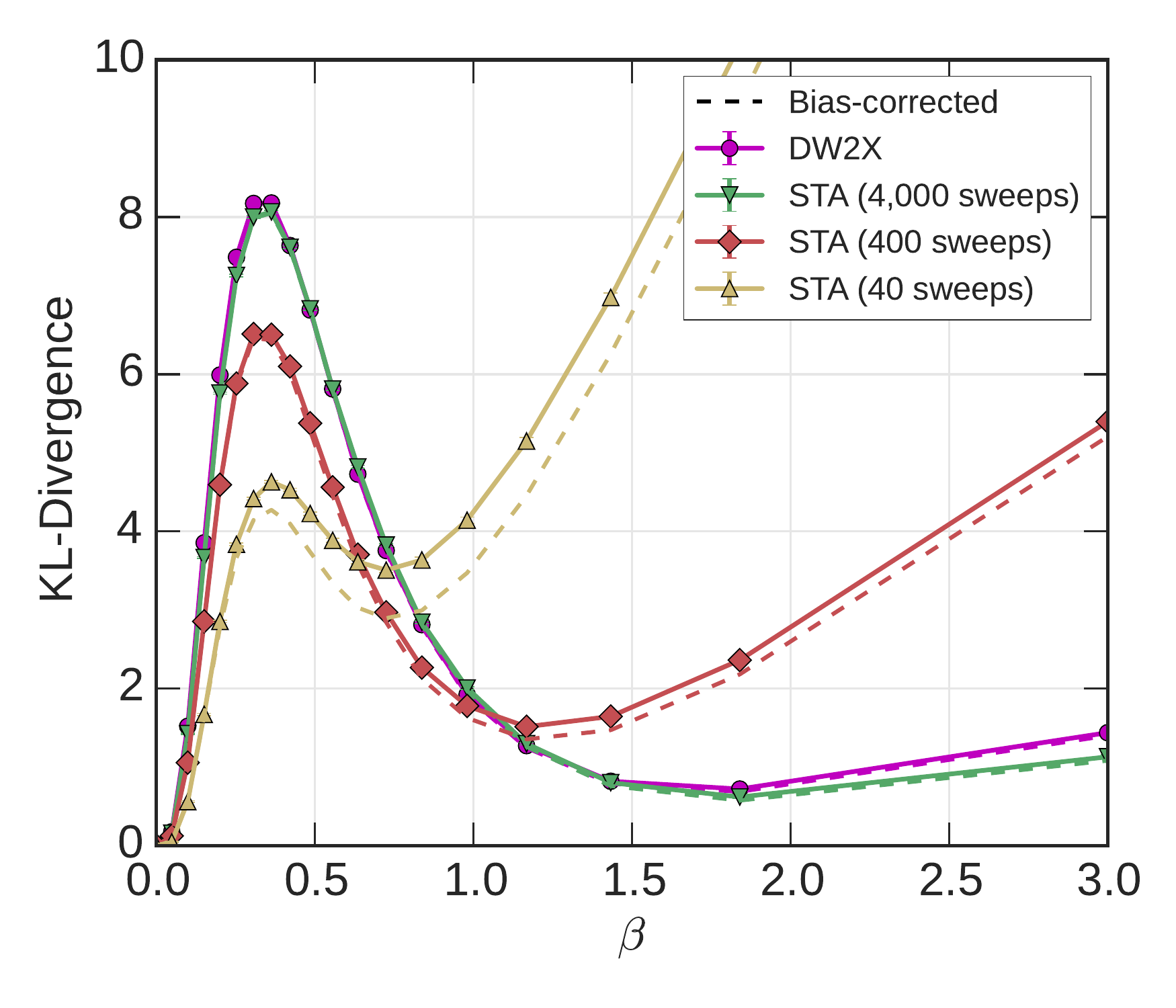}
  \caption{\label{fig:RAN1KLD} KL-divergence results for two exemplar instances of AC3 (left) and RAN1 (right) each at C4 scale (127 variables). Full lines indicate the estimate, and the dashed lines indicate jack-knife bias corrected estimates. The variance determined by the jack-knife method is negligible by comparison with symbol size. The bias is very large for 40-sweep annealing, indicating we have insufficient samples to properly determine the KL-divergence. Elsewhere we judge the bias not to significantly impact our conclusions.}
\end{figure}

The estimation of KL-divergence is described in supplementary materials, and allows us to measure KL-divergence post-processed RAN1 and AC3 problems, with reasonable precision up to C4 scale. In Figure \ref{fig:RAN1KLD} we demonstrate results for an exemplar on a C4 graph. The pattern observed is qualitatively similar to Figure \ref{fig:MSE2} in that we see a global minimum at $\beta=0$ reflecting the effectiveness of the post-processing, and a local minimum at intermediate $\beta$ that reflects a promising region for application of the annealer as a heuristic. As discussed in the supplementary materials, bias can be a problem with our KL-divergence estimator. For this reason we present a C4 (rather than C12) sized problem and only the post-processed (rather than unprocessed and post-processed) estimates. To assess the bias (the sensitivity of the estimate to the finite sample set size) the jack knife bias-corrected estimator is also shown~\cite{Efron:JK}. At C4 scale the bias does not significantly obscure the phenomena, particularly for the more powerful annealers (the STA with 4000 sweeps, and the DW2X).

\section{Discussion}
\label{sec:discussion}

In this paper we have considered several temperature estimators applied in the context of a physical quantum annealer set up for optimization (DW2X), and a comparable simply parameterized simulated thermal annealer (STA). We have demonstrated how different objective measures of closeness to the Boltzmann distribution respond differently to local and global distribution features. An important phenomenon we observe is that in annealed distributions we have a range of temperature estimates according to the method employed. We have shown that estimators indicating larger temperature are those responsive to macroscopic (global) features of the distribution. From an estimator perspective we have {\em global warming}: the more effective the estimator is in capturing global distribution features the higher the temperature that is typically indicated. Ergodicity breaking qualitatively explains the origin of this phenomenon in annealers, both the DW2X and STA. 
We have provided some practical guidelines in Section \ref{sec:Outline}.

Local distributional features, which are well characterized by a temperature, are easy to estimate by self-consistent methods. We showed in the main text standard methods for estimate temperature from samples drawn from a single distribution. Self-consistent approaches are efficient and approximate the target distribution from the same samples by which the heuristic is evaluated. We presented a simple form for this, but the principle generalizes. However, the main problem with such methods is that they may indicate a good fit on the basis of incomplete (or biased) information about the target distribution. If a heuristic sample set fails to see a representative set of modes in the distribution, then the evaluation will inevitably skewed by the missing information.

The local approximation method is able to capture an important difference between quantum and simulated annealers related to the difference between dynamics of the DW2X and STA. This is realized in the non-linear dependence of the local temperature estimate to variation of the DW2X rescaling parameter. Quantum simulations on single qubits provide a qualitatively accurate explanation for this phenomenon, and are described in supplementary material. 

Describing the global distribution in terms of temperature(s) is more tricky; we proposed KL-divergence and MSE as measures of deviation from the Boltzmann distribution, and based on these objectives developed estimators for the best temperature. Each of these objectives is affected in slightly different ways by deviations locally and globally from the Boltzmann distribution. The maximum likelihood estimator is more strongly affected by the local distribution than the minimum MSE estimator, and indicates a larger estimate of inverse temperature. To remove the local distribution effects we have proposed to take the initial distribution and apply local post-processing in order to isolate the macroscopic distribution effects that are truely a limitation on practical performance. Applying some degree of post-processing may also be valuable in practice, in particular for the DW2X since the post-processing is strictly classical and complementary to the quantum dynamics.

We emphasize that because efficient post-processing allows significant manipulation of the local temperature, we consider this temperature not particularly important in practical applications. If we post-process, the post-processing temperature itself will be synonymous with the local temperature; the local temperature need not be measured.

Important ideas incidental to the main thread are discussed in supplementary materials. These include: a description of single qubit experiments that explain the non-linearity of MLPL estimates in the DW2X; consideration of the effect of embedding on the distribution of samples; the development of an effective estimator for the Kullback-Leibler divergence; a consideration of spin-reversal transformations to mitigate sampling error in the DW2X; and an experiment to test how the choice of kernel (\ref{eq:BlocalApprox}) affects the locally self-consistent temperature estimates.

At various points in this paper we have included results both for the DW2X and STA. We have motivated a default parameterization for each algorithm in Section \ref{sec:DefaultParameterization}, and it should be clear that these choices are not aimed at, or appropriate for, a competitive comparison. It would be a complicated task to make a fair comparison, and it would also detract from the main theme of this paper since it would distort or disguise the phenomena we wish to highlight. We chose a default DW2X parameterization suitable for optimization, and two STA parameterizations that allow for qualitative comparison. It has been shown in experiments that both annealers can be improved with simple parameter modifications. It is interesting that, despite the fact that the DW2X annealer has been designed and tuned for optimization, it produces good statistics at intermediate temperature ranges, and that the STA with long anneal time shows a qualitatively similar behavior.

When considering the value of annealers in inference problems, it is also important not to forget a variety of other powerful inference methods that may achieve a similar objective. In particular, simple variations on the STA such as annealed (or population) importance sampling methods and other multi-canonical MCMC methods can often be tailored to the graphical structure of the problem under investigation~\cite{Neal:AIS,Hukushima:PA,Landau:GMC,hamze2004fields,Selby:sampling,Zhu:ECA}.

From the perspective of both errors on correlations and KL-divergence, the balance of evidence certainly indicates that there is potentially a sweet spot for application of either the DW2X or STA to sampling. This sweet spot may be problem type dependent, but can be tuned to a degree, by modification of the annealing parameters, and more importantly, by post-processing. However, evaluation of this sweet spot is difficult to do self-consistently, and someone interested in applications may have to undertake hard work to discover (and have confidence in) annealer performance. Having available curves such as those in Section \ref{sec:MLandMLPL}, probably for some weakly post-processed distribution, would allow parameters of the annealer to be set optimally. It may seem computationally intensive (defeating the value of the heuristic) to evaluate the macroscopic distribution before using an annealer, but it is reasonable to assume that for some classes of problems at large scale, the local and global temperature properties will be common across the class. In other time-dependent applications of annealers the statistics of the distributions being learned change slowly, so that only periodic evaluations of the temperatures may be required.

An important potential application of quantum annealers is in machine learning~\cite{Amin:QBM,PerdomoOrtiz:EET}, where other heuristic samplers (not annealers) are prevalent. A common heuristic used in machine learning is called contrastive divergence (CD)~\cite{hinton2002training,CP:ContrastiveDivergenceLearning}. $B_\beta$ is approximated in CD by taking the ground truths (a set of training examples) and evolving them by an MCMC procedure. This is an example of the post-processing scheme used and recommended in this paper, except that annealed samples are replaced by the ground truths. Like the annealing distribution, the distribution of training examples may be incorrect in both its local and global features, the effect of the MCMC procedure is to tidy up local distribution deviations. After post-processing the distribution is used directly - in effect the post-processing temperature is taken to be correct\footnote{In machine learning we can take the post-processing temperature to be $\beta=1$, without loss of generality.}, without consideration of potential macroscopic deviations. The success of this algorithm in practice indicates that learning procedures may be quite tolerant of macroscopic deviations from the Boltzmann distribution in application provided the local temperature is correct. This would be good news since it may be computationally expensive to quantify macroscopic deviations, but it is easy to measure and manipulate local temperature in annealers.

One feature of D-Wave quantum annealers that might lead us to consider a different approach is the quantum part, as already discussed. The single qubit freeze-out, and dynamical slow-down at larger scales, are described by quantum models. The quantum Boltzmann distribution may be a better fit to sample sets drawn from the DW2X, and perhaps in ``post-processing'' we should think of the quantum space as the target, rather than the classical one. This is certainly a promising direction for future work.

\section*{Acknowledgments}

The authors are grateful to Andrew King, Cathy McGeoch and Kevin Multani for their help in experimental design and method analysis. We also thank Alejandro Perdomo-Ortiz and John Realpe-G\'omez for input regarding their multi-canonical method.


\bibliographystyle{unsrt} 
\bibliography{Bibliography}
\appendix

\newpage
\section*{Figures}

\begin{figure}[htb]
  \centering
    \includegraphics[width=0.6\textwidth]{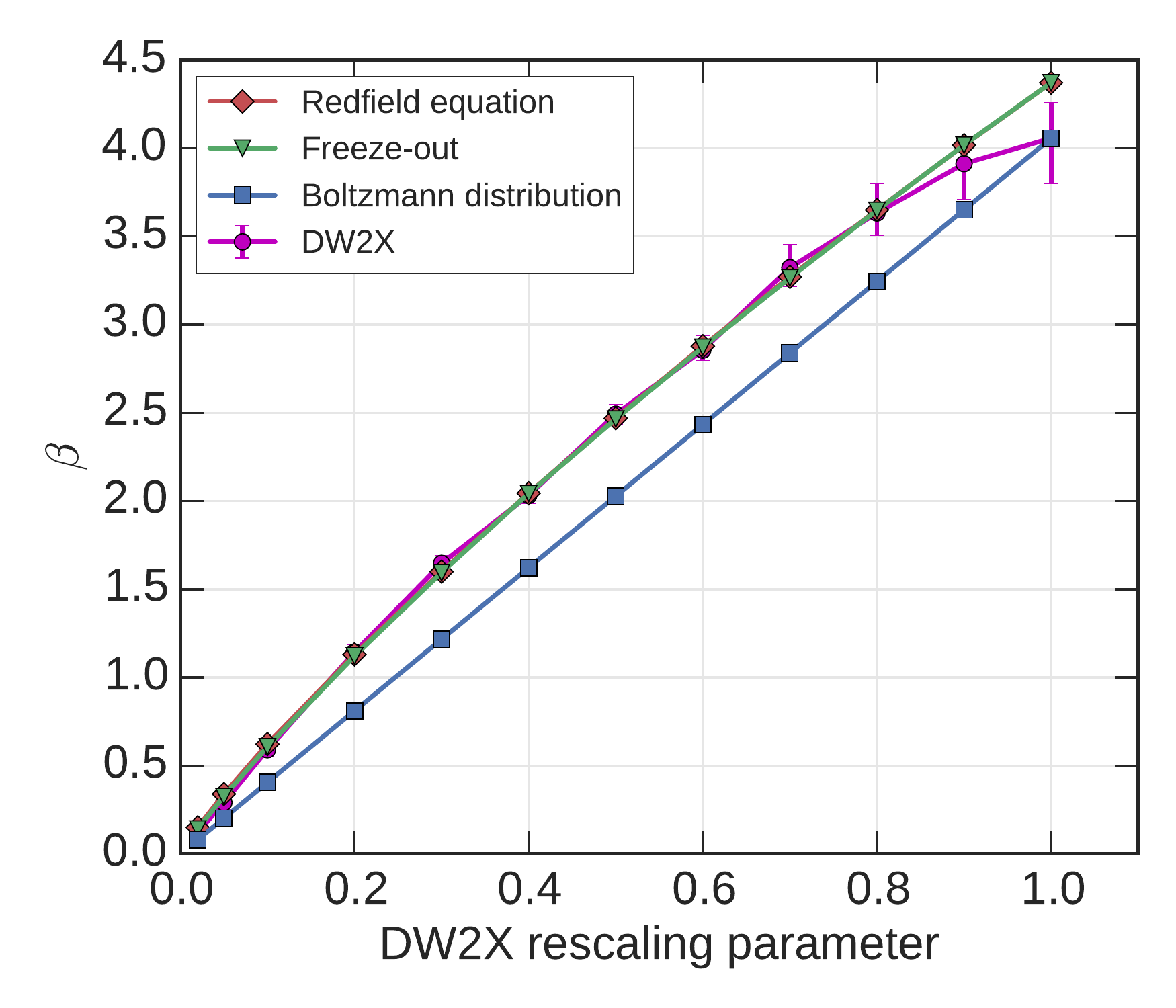}
\caption{\label{fig:rescalingDW2X1bit} An explanation for non-linearity of the MLPL estimator is possible through examination of single qubit dynamics in the DW2X. This study is described in supplementary materials. MLPL becomes equivalent to maximum likelihood in the limit of a single qubit. The single qubit $\beta$ dependence on the rescaling parameter, for the single qubit Hamiltonian $H(x)=x_i$ is shown. We plot the DW2X median value together with 25-75 quantiles as error bars. Simulation of the physical dynamics gives the Redfield curve, and the ``freeze-out'' curve shows is based on the assumption of a single freeze-out point. The theory thus is in agreement with experiment. The DW2X is not equilibrated at the end of the anneal even for a single qubit model. By contrast, in a system that is locally or globally equilibrated at the end of the anneal a linear dependence would be expected, as is seen for the STA. }
\end{figure}

\begin{figure}[htb]
  \centering
    \includegraphics[width=0.45\textwidth]{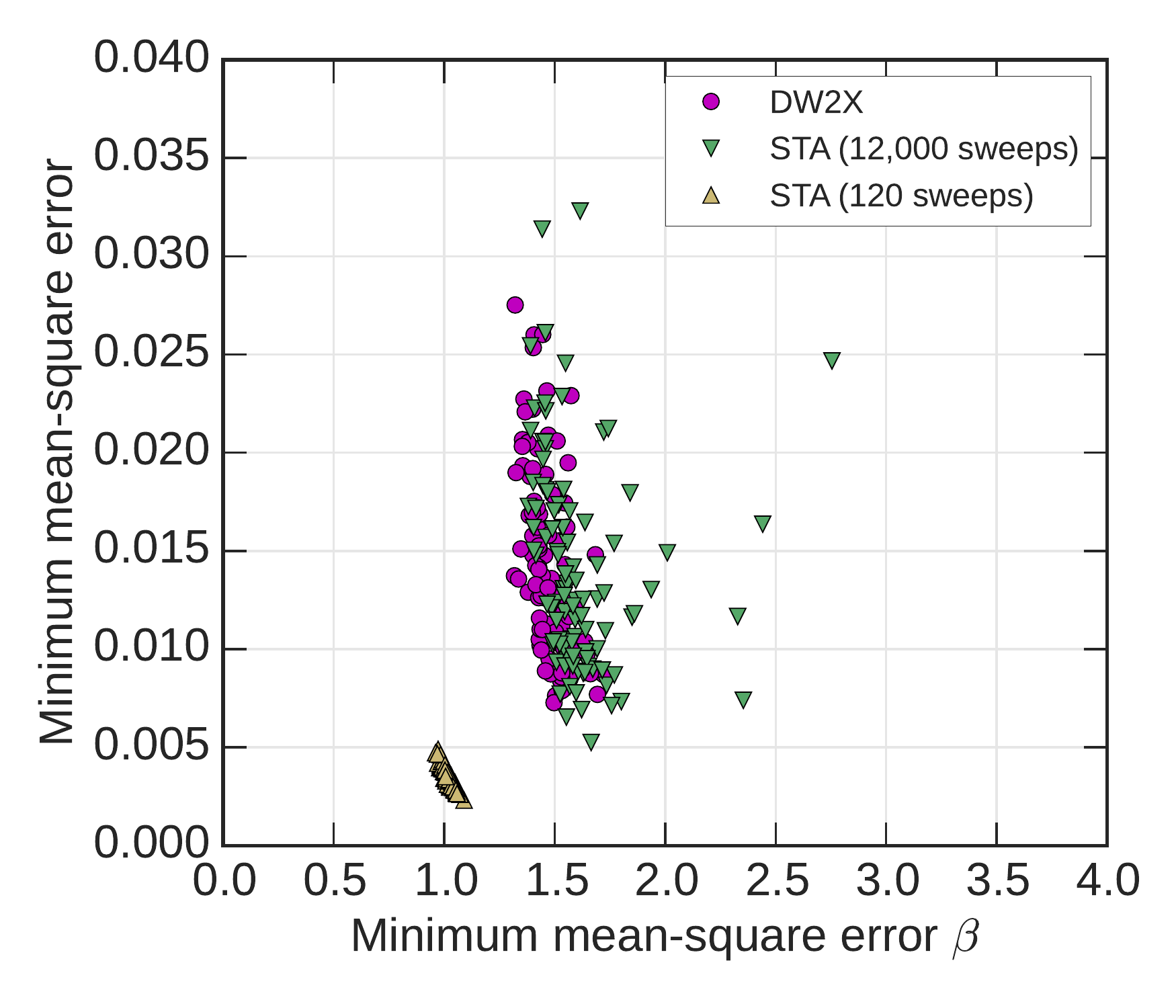}
    \includegraphics[width=0.45\textwidth]{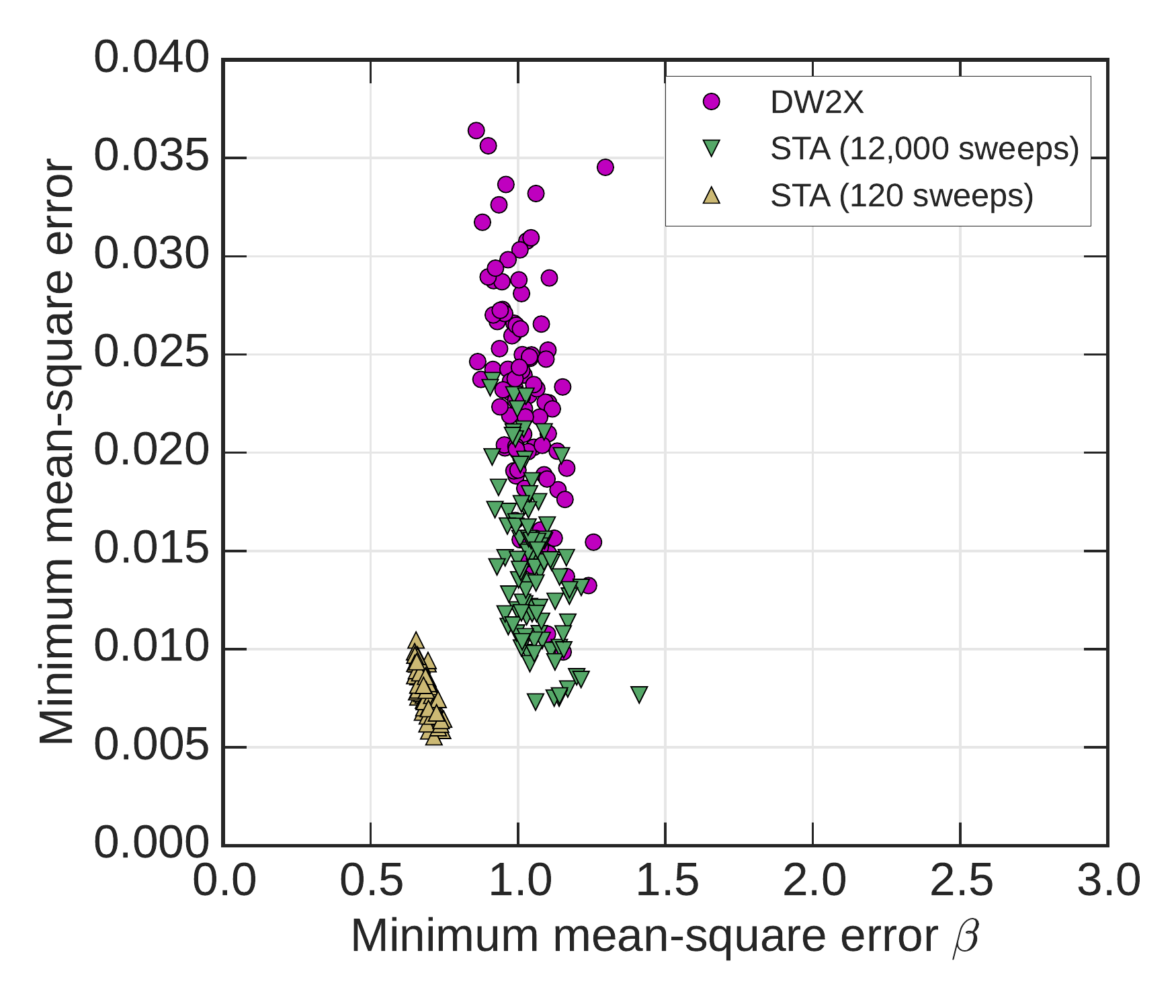}
    \caption{\label{fig:EstPerEst2_PP} As per Figure \ref{fig:EstPerEst2}, but with distributions modified by one sweep of blocked Gibbs sampling. Objectives are improved everywhere very significantly, and by a comparable fraction across the different annealers.}
\end{figure}

\end{document}